\documentclass[a4paper,11pt]{article}
\usepackage{jcappub}
\usepackage{amsmath}
\usepackage{enumerate}
\usepackage{natbib}
\usepackage{url}
\usepackage{xcolor}
\usepackage{comment}
\usepackage{orcidlink}

\defcitealias{B21}{B21}
\defcitealias{DF22}{DF22}
\defcitealias{FZH04}{FZH04}
\defcitealias{ZH06}{ZH06}

\newcommand{\appropto}{\mathrel{\vcenter{
  \offinterlineskip\halign{\hfil$##$\cr
    \propto\cr\noalign{\kern2pt}\sim\cr\noalign{\kern-2pt}}}}}

\title{The effect of a short mean free path on HII regions and 21-cm tomography during reionization}

\author[a]{Michael M. Wyatt\orcidlink{0009-0008-7752-7800},}

\author[a]{Steven R. Furlanetto\orcidlink{0000-0002-0658-1243},}

\author[b,a]{and Mary H. Minasyan\orcidlink{0000-0003-3393-2819}}

\affiliation[a]{Department of Physics and Astronomy, University of California, \\
Los Angeles, CA 90095, USA}

\affiliation[b]{Department of Physics, California Institute of Technology, \\ Pasadena, CA 91125, USA}

\emailAdd{michael.wyatt@physics.ucla.edu}

\abstract{Recent measurements of the mean free path (MFP) of ionizing photons at $z=6$ find that it is significantly shorter than extrapolations from lower redshifts. This has a substantial impact on the topology of reionization and thus the prospects of tomography of the 21-cm signal from upcoming radio interferometers. In this work we develop the first analytic model of reionization which explicitly incorporates the MFP as a free parameter, allowing us to transparently explore its effect on the process. Our model is based on the excursion set formalism with an ionization condition which accounts for absorptions parameterized through the MFP. With the goal of direct observational comparison, we also include additional modifications which make our model particularly suitable for predicting one-point statistics of the ionization field (and 21-cm signal), which are among the fundamental quantities for tomography. We find that the effect of the MFP is much more significant during the later stages of reionization, and that including a shorter MFP reduces the size of H{\footnotesize II} regions by around an order of magnitude towards the end of reionization compared with analytic models which do not account for the MFP. We find that the reported MFP value produces a contrast in the 21-cm signal of $\mathcal{O}$(1 mK) or less at resolutions $\theta \sim $ 15--35 arcmin, an order of magnitude below naive estimates and up to a factor of several smaller than when using a larger MFP value extrapolated from low redshift, requiring significantly more sensitivity for imaging. We compare the contrast to noise estimates for arrays similar in size to HERA and the first phase SKA-Low and find that SKA has sufficient sensitivity for direct imaging (at the largest scales considered), while the predicted signal will be challenging for arrays similar in size to HERA. Our model indicates that more detailed sensitivity estimates are warranted in the context of a shorter MFP.}

\begin{document}
\maketitle
\flushbottom

\section{Introduction} \label{sec:intro}

Understanding reionization remains one of the major goals of modern cosmology as it links the early Universe on an enormous range of scales, from sub-kpc scales relevant to galaxies, black holes, and star formation, to hundreds of Mpc scales where features in the ionization field are present, promising to elucidate many processes in early galaxy formation and evolution. In particular, constraints on the evolution of the topology of the ionization field, either through direct observations such as those of the 21-cm signal (e.g., \cite{Pober14,Greig15,Liu16,Kern17,HERAI_constraints,HERAI_improved,Nunhokee25,Mertens25,Munshi25}; for reviews see \cite{FOB06,Pritchard12,Liu20}) or indirect observations such as through detections of Lyman-$\alpha$ emitters (LAEs; e.g., \cite{McQuinn07_LAE,Trapp23,Tang24,Witstok25,Umeda25_Silverrush,Mason25,Kageura25}), allow constraints on properties of the sources driving reionization through comparisons to theoretical models. However, the large dynamic ranges relevant to reionization also make it difficult to simulate efficiently; although great strides have been made in recent years (e.g., \cite{CROCI, CROCII, CoDaIII, THESANa, THESANb, THESANc}), fully numerical simulations of the process remain computationally costly. Because the details of reionization are relatively unconstrained, efficient simulations which can probe wide swaths of parameter space remain an integral part of the exploration of this epoch.

One of the most important effects dictating the evolution of the ionization field is the absorption of ionizing photons due to intervening neutral gas in the intergalactic medium (IGM). In many semi-numerical and analytic models, this is accounted for only approximately through a maximum scale considered in the excursion set $R_{\rm max}$, essentially setting a strict upper limit on the distance that ionizing photons are allowed to travel (e.g., \cite{21cmFAST, Zahn11, Alvarez12, Majumdar14}). However, when averaged over all lines of sight from a source, the attenuation of ionizing photons due to neutral gas in the IGM is better described through an exponential decline with distance~$\propto~{\rm e}^{-r/\lambda}$, where $\lambda$ is the mean free path (MFP) of these photons. Thus, the traditional approach of photon attenuation in many semi-numerical and analytic models serves to underestimate the attenuation of photons on scales $< R_{\rm max}$ and oversuppresses their ability to travel distances $>R_{\rm max}$. The MFP is set by a combination of both the topology (i.e., ionized bubble sizes), which dictates the distance that photons travel from sources before encountering the neutral IGM, as well as dense, neutral clumps in the already-ionized IGM in the form of e.g., damped Lyman-$\alpha$ and Lyman limit systems.

A recent measurement of the MFP found $\lambda = 5.25^{+4.55}_{-3.15}$ cMpc at $z=6$ (\cite{B21}; hereafter \citetalias{B21}), significantly shorter than extrapolations from lower redshifts and results from many simulations (e.g., \cite{D'Aloisio20, Keating20}). Although the physical origin of the short MFP is not clear \citep{Cain21}, since these measurements were made, some simulations have successfully reproduced the shorter value \cite{THESANb, CoDaIII}. Recent follow-up studies have also found small mean free paths \cite{Zhu23, Satyavolu24}, though the latter found a somewhat larger mean best-fit value of $10.43$ cMpc. Such MFP lengths are at or below the characteristic scales of ionized bubbles found in existing models during the latter half of reionization \citep{FZH04, McQuinn07, Mesinger07, Zahn11, Lin2016, Neyer24}, meaning that this change should have a non-negligible impact on the topology of the ionization field, and as such demands a more realistic treatment of the MFP in semi-numerical and analytic models. There is one recent example of an explicit implementation of the MFP in a modification to the \texttt{21cmFAST} semi-numerical code \citep{DF22} (hereafter \citetalias{DF22}) which was also recently adopted into the full public release of the code \citep{Davies25}, however as of yet there have been no analytic models which do the same. In this work, we develop an analytic model of reionization which explicitly incorporates the value of the MFP as a free parameter, allowing us to transparently isolate and explore its effect on the topology of reionization.

We note that there have also been analytic and semi-numerical efforts to model the absorption of ionizing photons through methods other than the direct inclusion of an MFP. For example, \cite{sobacchi_18} developed a prescription for tracking inhomogeneous recombinations in the IGM in a semi-numerical model (\texttt{21cmFAST}), which naturally results in an MFP for the ionizing photons. Using this work as a basis, \cite{Qin25} constrained their model using observations of the high-redshift Lyman-alpha forest. Given a prescription for the absorption of photons due to Lyman limit systems, the authors were able to recover a total MFP (i.e., one that is also dependent on the topology) at $z=6$ that is consistent with recent measurements. Based on \texttt{21cmFAST}, \cite{Xu17} developed a semi-numerical method which models the end stages of reionization using an ionizing background that is computed by accounting for photon absorption. In \cite{Zhu23_Island} the authors developed this model further by including inhomogeneous recombinations in a similar manner to \cite{sobacchi_18}. \cite{Furlanetto05} also incorporated recombinations in an analytic model --- we return to a more detailed comparison with this model in section \ref{S:semi-ana}. As the authors of \citetalias{DF22} pointed out, methods such as \cite{sobacchi_18} still rely on a particular sub-grid model for density fluctuations, while in reality the physics dictating the MFP may be far more complex \cite{Park16, D'Aloisio20}. An explicit implementation, especially when combined with a simple, efficient analytic model allows us to more easily isolate and directly probe the effects of the MFP length on the evolution of the ionization field, free from any physical assumptions of the IGM, and such a simple approach can also be more easily included in parameter inference.

Any change to the topology of reionization should also affect various observables. Efforts to produce directly observable quantities from analytic and semi-numerical models are often focused on global quantities such as the global ionized fraction or power spectrum. These are attractive because they are relatively simple quantities which nevertheless provide powerful constraints. However, while sufficient to completely describe the statistics of a Gaussian random field, the power spectrum provides incomplete information about the non-Gaussian ionization field. Because of this, tomography (direct imaging) of the 21-cm signal remains the ultimate goal for observations of reionization. Among the most fundamental quantities related to imaging are one-point statistics, i.e., the probability distribution function (PDF) of the 21-cm brightness temperature on a given scale, and its higher-order moments (such as variance, skewness, and kurtosis). This set of statistics has been explored analytically \citep{FZH04} and numerically \cite{Mellema06,Wyithe07,Ichikawa10,21cmFAST,Watkinson14,Watkinson15,Kubota16}, and in the context of observational feasibility \cite{Kittiwisit18,Kittiwisit22, Kim25}, and other more  complex but related quantities have also been explored \citep{Barkana08}. For these reasons, we develop our model with the specific intention of producing the one-point PDF of both the ionization field and the 21-cm brightness temperature. Compared with previous analytic explorations of this statistic, our approach improves the treatment of substructure and incorporates a shorter MFP for the first time.

With telescopes such as the Hydrogen Epoch of Reionization Array (HERA; \cite{DeBoer17}) operational \citep{HERAI, HERAII} and with efforts underway for direct imaging \citep{Xu22, Kim25}, and the Square Kilometre Array (SKA; \cite{Koopmans15}) currently under construction with early operations possible by the end of the decade, now is the opportune time to explore the implications of the MFP on the prospects for mapping the 21-cm signal. The one-point PDF also has applications beyond imaging of the ionization field itself. For example, \cite{Mesinger08} showed that the one-point PDF of visible LAEs (which is modulated by the PDF of the ionization field) can be used as an effective constraint of reionization, as well.

This paper is structured as follows: in section \ref{S:semi-ana} we describe the modifications we make to the standard analytic excursion set model to better account for the MFP. Subsequently, in section \ref{s:growth-of-HII} we discuss implications for the sizes of HII regions. In section \ref{s:one-point-PDF-description} we describe further modifications we make to better address substructure and partial ionizations, and thus compute the one-point PDF. In section \ref{s:one-point-PDF} we show the dependence of the one-point PDF on various parameters within the model. In section \ref{s:sensitivity} we compute the typical contrast of the 21-cm signal and compare this to estimates for the noise levels of telescopes of similar size to HERA and SKA-Low. Finally, in section \ref{s:conclusion} we conclude.

Throughout this work we use a flat $\Lambda$CDM cosmology with $\Omega_{\rm m} = 0.3111$, $\Omega_{\Lambda} = 0.6889$, $\Omega_{\rm b} = 0.0489$, $\sigma_8 = 0.8102$, $n_s = 0.9665$, and $h = 0.6766$, consistent with the results of \cite{planck_planck_2020}.

\section{Mean free path--dependent ionization criterion}\label{S:semi-ana}
Our model builds off of the so-called `photon-counting' argument originally presented in \cite{FZH04} (hereafter \citetalias{FZH04}) but with important modifications which allow it to more realistically account for the mean free path of ionizing photons. The original photon-counting model of reionization relates the ionized fraction of a region to the collapse fraction and the ionizing efficiency. The collapse fraction $f_{\rm{coll}}$ is the fraction of baryons in a region which are within collapsed halos above some minimum halo mass $M_{\rm{min}}$ for star formation, and the ionizing efficiency $\zeta_0$ is the number of ionizing photons emitted per collapsed baryon. A region is said to be fully ionized when $1 = \zeta_0 f_{\rm{coll}}$. Thus, such methods `count' the number of ionizing photons produced by collapsed objects within the region and compare this to the total number of baryons in order to determine if there are enough photons to ionize the entire region, assuming that each photon is able to ionize one hydrogen atom.\footnote{In more detail, $\zeta_0$ also accounts for the average \textit{cumulative} number of recombinations; a constant factor which represents the number of photons needed to ionize each hydrogen atom (due to the number of times the hydrogen recombines) is absorbed into $\zeta_0$.\label{fn:zeta}}

Photon-counting methods rely on the mathematics of the excursion set formalism (see \cite{Z07} for a comprehensive review) in order to solve for the size distribution of ionized regions. The collapse fraction of a region with radius $R$ and linear density $\delta$ is given by $f_{\rm{coll}}(\delta|S)~=~\mbox{erfc} [ ( \delta_c - \delta) / \sqrt{2 (S_{\rm min} - S)} ]$, where $S = \sigma^2(R)$ is the variance of the linear density field smoothed over a radius $R$ and can be computed by assuming that the density fluctuations are a Gaussian random variable, $\delta_c$ is the critical minimum density for a region to virialize, and $S_{\rm min}$ is the variance corresponding to the minimum mass to form a collapsed object, $M_{\rm min}$, which we take to be $10^8$ M$_\odot$.\footnote{It is common to refer to the smoothing scale by the variance, the corresponding length, or a mass, where the mass is given assuming the region is at the mean density.\label{fn:mass-radius}} It is important to note that $S$ is a monotonically \emph{decreasing} function of $R$. As the collapse fraction is a monotonic function of density, the ionization criterion $1 = \zeta_0 f_{\rm coll}(\delta|S)$ can be inverted for the threshold (minimum) density required for a region to be fully ionized as a function of radius $R$ (or equivalently as a function of scale $S$). When performed in equal steps in $S$, the density field smoothed over progressive scales can be approximated as an uncorrelated random walk. Taking the threshold ionizing density to be an absorbing barrier, this allows one to solve for the first-crossing distribution of such a barrier analytically. Then, the largest scale at which the density crosses the barrier is identified as an ionized region and the sizes of ionized bubbles can be described in a statistical sense.

Many semi-numerical models essentially recreate the excursion set method within a simulation box of the density field; beginning on a large radius, the density is averaged over progressively smaller scales until the density crosses the threshold for ionization --- at that point, the entire region is considered ionized. If the density never crosses the threshold prior to reaching a minimum radius (generally the scale of an individual cell within the simulation box), then the cell is partially ionized using $Q = \zeta_0 f_{\rm coll}$. Some models, such as \texttt{21cmFAST}, choose for computational efficiency to ionize only the \textit{central} cell of the ionized region when the threshold density is reached --- a method often referred to as `central pixel flagging' \citep{21cmFAST}. \citetalias{DF22} identified that central pixel flagging is equivalent to integrating the ionizing flux from surrounding sources onto the central pixel. Given this equivalence and using the central pixel flagging method, the authors derived a new ionization criterion for excursion set--based models which accounts for the attenuation of ionizing photons through a constant mean free path, and applied this to a semi-numerical reionization model. For completeness, we repeat the derivation here.

We begin by considering a region of comoving volume $\Delta V$ which is emitting ionizing photons. This volume is a proper distance $r_p$ away from a cell of proper volume $\Delta r_p^3$, where the proper length is related to the comoving length through $r_p = r/(1+z)$. The flux of photons reaching the cell from the emitting volume is

\begin{equation}
    f = \frac{\dot{N}_{\rm ion}}{4 \pi r^2_p},
\end{equation}
where $\dot{N}_{\rm ion} = \epsilon \Delta V$ is the rate at which ionizing photons are produced, and $\epsilon$ is the emissivity of ionizing photons. The rate at which photons enter the cell $\dot{N}$ is thus the flux multiplied by the surface area of the region:

\begin{equation}
    \dot{N} = \frac{\epsilon \Delta V}{4 \pi r^2_p} \Delta r_p^2 = \frac{\epsilon \Delta V}{4 \pi r^2} \Delta r^2,
\end{equation}
where in the last step the common factors of redshift cancel. If we then imagine that the volume emitting ionizing photons is a spherical shell of radius $r$ and thickness $dr$ (and thus $\Delta V = 4 \pi r^2 d r$), and that the cell is placed in the center of this shell, we are left with $\dot{N}_{\rm sh} = \epsilon \Delta r^2 dr$. Integrating this rate over the history of the Universe leaves us with

\begin{equation}
    N_{\rm sh} = \zeta_0 f_{\rm coll} n_b \Delta r^2 dr,
\end{equation}
where $n_b$ is the baryon density inside the shell. Next, we account for the mean free path by acknowledging that on average only a fraction ${\rm e}^{-r/\lambda}$ of these photons will reach the central cell, while the others will be absorbed by intervening neutral gas, leaving us with $N_{\rm sh} = \zeta_0 f_{\rm coll} n_b \Delta r^2 {\rm e}^{-r/\lambda} dr$. If $N_{\rm tot}$ is the total number of ionizing photons reaching the central cell from a sphere of a radius $R$ ignoring absorption, and $N_{\rm tot}^\lambda$ is the same quantity accounting for absorption, then

\begin{equation}
    \frac{N_{\rm tot}^\lambda}{N_{\rm tot}} = \frac{\int_0^R \zeta_0 f_{\rm coll} n_b \Delta r^2 {\rm e}^{-r/\lambda}  dr}{\int_0^R \zeta_0 f_{\rm coll} n_b \Delta r^2 dr} = \frac{\lambda (1 - {\rm e}^{-R/\lambda})}{R}
\end{equation}
is the factor which describes the `excess' photons required to ionize the central cell given absorption compared with those required without accounting for absorption. Simply including this factor in the original criterion gives us a new ionization criterion which accounts for the average photon absorption through an MFP $\lambda$, leading to the result of \citetalias{DF22}:

\begin{equation} \label{eq:MFP-ion-criterion}
    1 = \frac{\lambda ( 1-{\rm e}^{-R/\lambda} )}{R} \zeta_0 f_{\rm{coll}}.
\end{equation}

We note that there is some nuance in the way that the MFP is defined in this case. Throughout reionization, ionizing photons can be divided into two categories: those which are ionizing neutral regions of the IGM and thus growing ionized bubbles (such photons are absorbed at the edges of ionized bubbles) and those which are absorbed by dense clumps --- such as Lyman limit systems --- in already ionized space (such photons are absorbed \emph{within} ionized bubbles). These two regimes have different MFP values, and the total MFP is determined by both through: $\lambda_{\rm tot}^{-1} = \lambda_{\rm LLS}^{-1} + \lambda_{\rm bub}^{-1}$ (e.g., \cite{Alvarez12}). As reionization progresses, one value should dominate over the other; early in reionization, small bubble sizes limit the distance that photons can travel, while the end of reionization is expected to result in a sharp transition in the MFP as ionized bubbles give way to a nearly fully ionized IGM and $\lambda_{\rm LLS}$ dominates. This transition is seen in \citetalias{B21} and the value measured at $z=6$ has indeed been interpreted as the total MFP length including the effects of both topology and IGM clumpiness \citep{Roth24}. 

Importantly, the MFP measured by \citetalias{B21} is spatially averaged, but instantaneous with respect to time, while the criterion given by eq. \ref{eq:MFP-ion-criterion} is averaged both spatially and in time. As the criterion assumes that a region of space requires on average a factor $\lambda ( 1-{\rm e}^{-R/\lambda} )/R$ more photons to be ionized compared with one within a Universe with an infinite MFP, it would be appropriate to use this average MFP length. However, reionization is a process driven by spatial inhomogeneities, and such an MFP value would be too short within larger ionized bubbles, artificially reducing their size. On the other hand, because the criterion relies on the \emph{cumulative} number of ionizing photons, $\lambda$ in eq. \ref{eq:MFP-ion-criterion} is also in a sense a time-averaged value. In this case, the value measured at $z=6$ is larger than is appropriate for the condition, as smaller bubbles earlier in reionization will result in a shorter MFP. Given these opposing tendencies, the approximations of the ionization criterion, and other significant sources of uncertainty, we consider the values measured by \citetalias{B21} to be an appropriate point of comparison for our model.

\begin{figure*}
\centering
    \centering
    \includegraphics[width=0.5\linewidth]{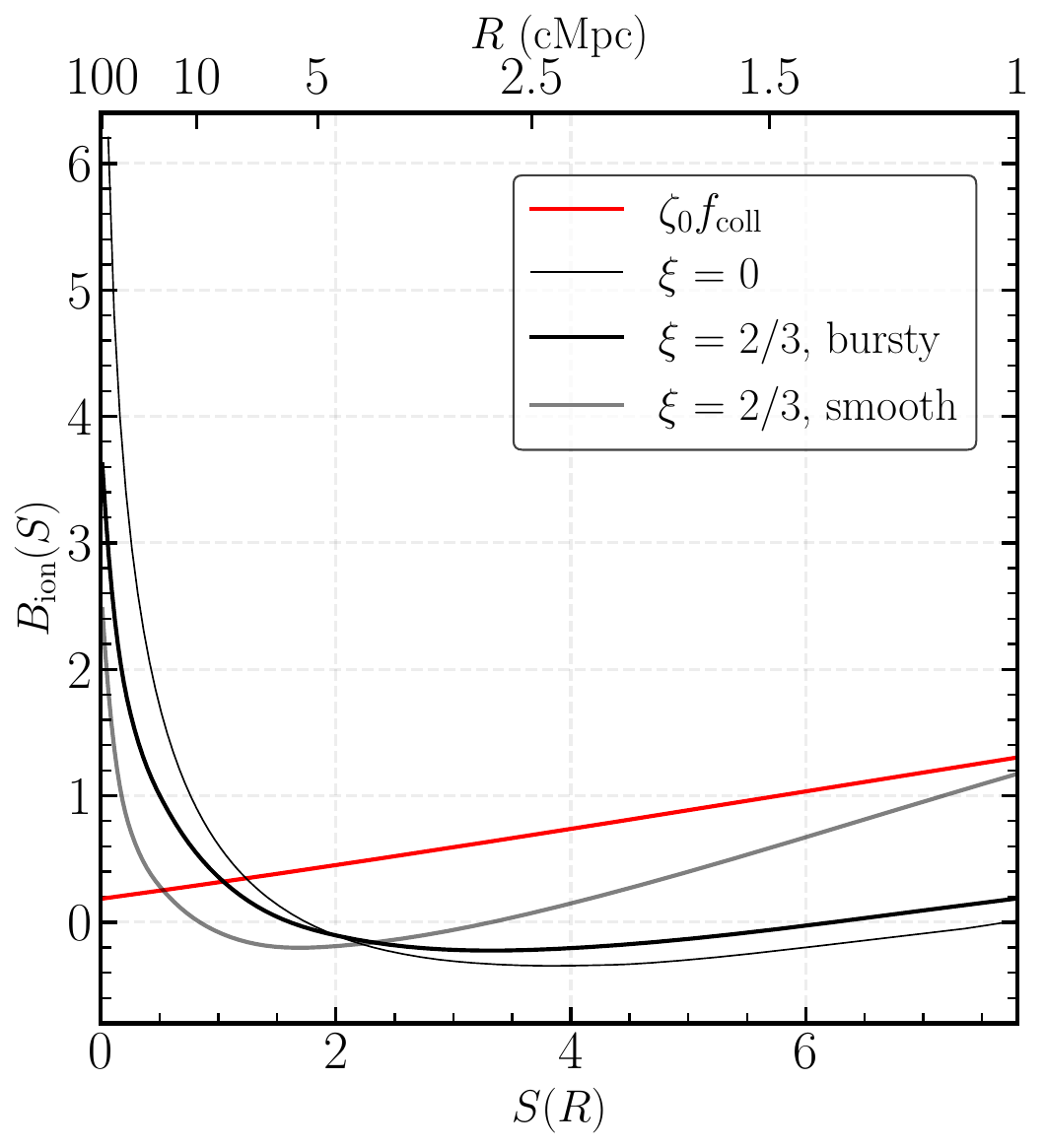}
\caption{\textbf{A short MFP limits the sizes of fully ionized regions by enforcing larger density requirements for a fully ionized region at larger radii.} Lines show the barrier used in the excursion set formalism for the case where no MFP is included (red line) and for the ionization criterion which includes the MFP (black lines). The thin dark line assumes a constant ionizing efficiency with no mass dependence ($\xi=0$), while the two thicker lines assume a power law on the mass-dependence of $\xi = 2/3$ which reflects energy-regulated stellar feedback (see section \ref{ss:mass-depend}). The lighter of these two lines assumes a smooth star formation rate across the entire halo mass range, while the darker line assumes bursty star formation. All cases use $\lambda = 5$ cMpc and are normalized to have the same global neutral fraction of $x_{\rm HI} = 0.1$ at $z=6$.}
    \label{fig:mass_dep_zeta_barriers}
\end{figure*}

The new criterion can be applied to an excursion set model in precisely the same way as the old photon-counting criterion; inverting the relationship for the density of the region gives the scale-dependent threshold density required for a region to be fully ionized:
\begin{equation} \label{eq:ion-bar}
    B_{\rm{ion}}(S) = \delta_c(z) - \sqrt{2(S_{\rm{min}}-S)} \, {\rm{erfc}}^{-1}\left[ \frac{R(S)}{\left( 1 - {\rm{e}}^{-R(S)/\lambda} \right) \zeta_0 \lambda} \right].
\end{equation}
As before, if the minimum radius in the excursion set is reached without the criterion having been met on a larger radius, $Q = \zeta_0 f_{\rm coll}$ is used to partially ionize the region; so long as the minimum scale is sufficiently small compared to the MFP, one can reasonably ignore the effect of the MFP on such scales. An example of this barrier compared with that of the older photon-counting models is shown in figure \ref{fig:mass_dep_zeta_barriers}. The smoothing scale is shown both in terms of $R$ and the corresponding $S$. Compared with the more traditional analytic barrier, a steep upturn at large radii limits the size of fully ionized regions (an asymptotic behavior actually sets a strict maximum to the size of an ionized bubble with this barrier), which is the behavior we expect from a small MFP. 

A qualitatively similar barrier was also found in \cite{Furlanetto05}. In this work, the authors accounted for recombinations in an analytic model by balancing the \textit{rate} of photon production with the recombination rate, as opposed to the integrated \textit{number} of photons with hydrogen atoms as is done with eq. \ref{eq:MFP-ion-criterion} (or the simpler $1 = \zeta_0 f_{\rm coll}$), where we note again that the latter accounts for the cumulative number of recombinations through $\zeta_0$ (see footnote \ref{fn:zeta}). As the authors in \cite{Furlanetto05} pointed out, these approaches represent two limiting cases: in reality sources must both have produced enough photons to ionize the entire region plus be producing enough at a given time to counteract recombinations. The approach in \cite{Furlanetto05} indirectly sets the MFP length by enforcing that the MFP in a given ionized bubble must be greater than or equal to the bubble radius. As in the case of enforcing a strict upper limit to the excursion set (see section \ref{sec:intro}), this approach artificially limits the distance that photons can travel. Indeed, with the goal of a tractable analytic model, the authors of \cite{Furlanetto05} made this very approximation and only allowed fully ionzied regions below some $R_{\rm max}$. While qualitatively similar to our own barrier, this solution represents a more extreme approximation.

\subsection{21cmFAST simulation boxes}\label{ss:21cmFASTcompare}

\begin{figure}
    \centering
    \includegraphics[width=0.88\linewidth]{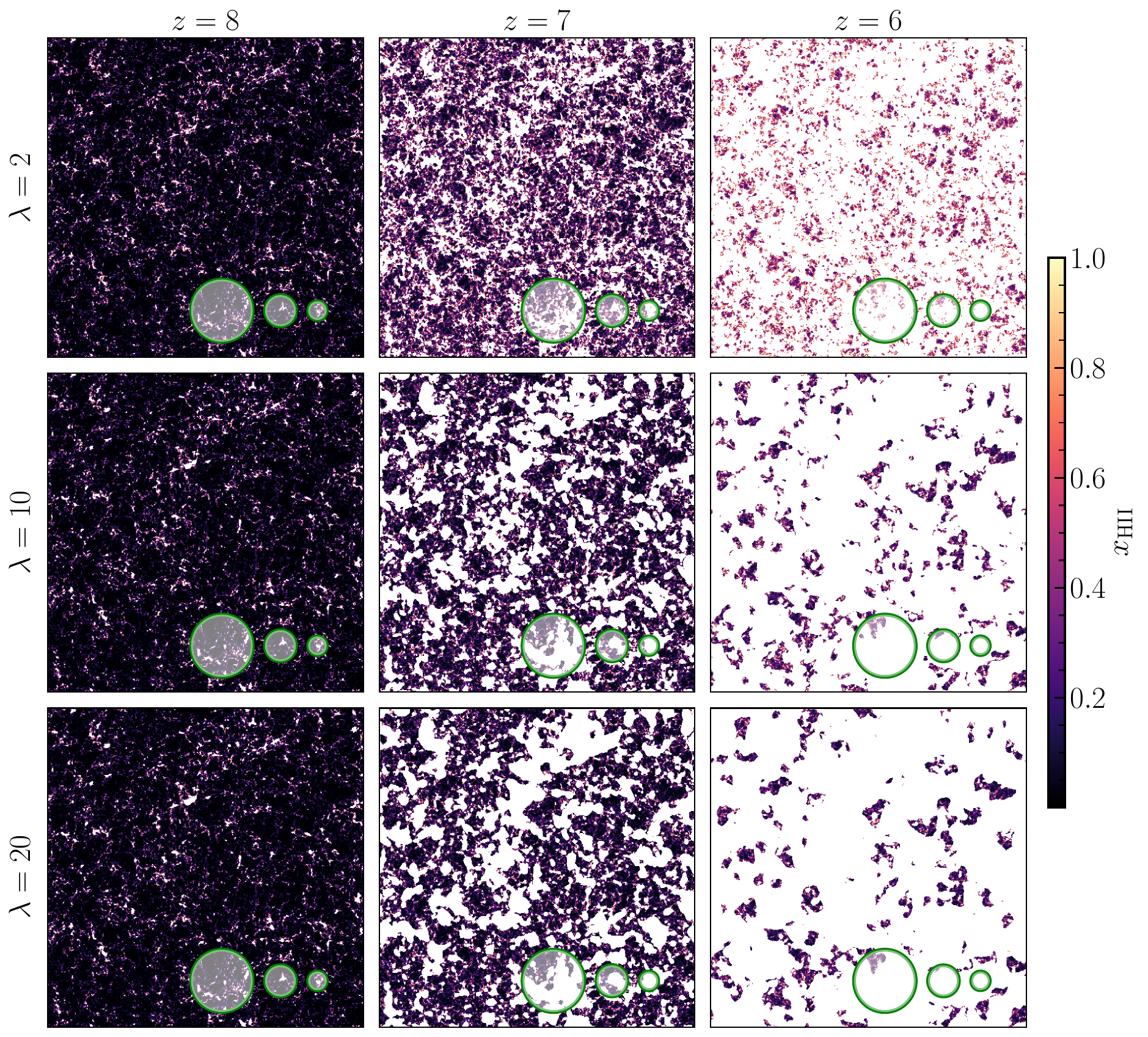}
    \caption{\textbf{A smaller MFP value results in a significant change in the ionization field.} Each panel shows a slice from an ionization box output by \texttt{21cmFASTv1} using the ionization criterion which accounts for the MFP, where each box is 512~cMpc on a side and 1 cMpc thick. Columns show different redshifts ($z = 8,7,6$) at fixed global neutral fractions ($\left< x_{\rm HI} \right> = 0.9, 0.5, 0.1$) and rows show different values of the MFP ($\lambda = 2,10,20$~cMpc), corresponding to the limits measured by \citetalias{B21} (2 and 10 cMpc) and the value used in many of the results in \citetalias{DF22} (20 cMpc), before the shorter MFP value was reported. The colorbar shows the ionized fraction of each pixel, where white pixels are fully ionized. For comparison, green circles are shown with radii $50,25$, and 15 cMpc, corresponding to smoothing scales used throughout this work. Slices of the ionization field clearly show several trends. Firstly, at the lowest global neutral fraction, smaller MFP values result in a striking difference in the ionization field, with neutral islands appearing smaller and more numerous and with less extreme neutral fractions, resulting in a more homogeneous ionization field. Secondly, the MFP has a much more significant effect later in reionization; panels at $z=8$ ($\left< x_{\rm HI} \right> = 0.9$) show very little difference between MFP values, while differences are visually clear by $\left< x_{\rm HI} \right> = 0.5$ and particularly by $\left< x_{\rm HI} \right> = 0.1$. In all cases the ionizing efficiency is assumed to be constant with mass ($\xi=0$; see section \ref{ss:mass-depend}).}
    \label{fig:21cmFASTslices}
\end{figure}

\citetalias{DF22} applied the new criterion to \texttt{21cmFASTv1}, and although they provided visual comparisons of the ionization field for various MFP lengths, their work was completed prior to more recent measurements of a shorter MFP and as such the values considered were significantly larger than those measured in \citetalias{B21}. As the change in the ionization boxes for different MFP lengths is visually apparent, it is illustrative to compare them using smaller MFP values in order to conceptualize the changes to other quantities explored in this work and to further motivate the exploration of the effect of the shorter MFP. In figure \ref{fig:21cmFASTslices} we show 2D slices of the ionization field output by \texttt{21cmFASTv1} using the ionization criterion of eq. \ref{eq:MFP-ion-criterion} at various global neutral fractions. We show results using $\lambda = 2$ and $10$ cMpc, bracketing the limits reported by \citetalias{B21}, and a larger value of 20 cMpc, the value used in the ionization boxes shown in \citetalias{DF22}. There is a stark difference in the ionization field for $\lambda = 2$ cMpc compared with $\lambda = 10$ or 20 cMpc. Keeping in mind that each case is normalized to the same global ionized fraction, for shorter MFP values neutral islands are clearly smaller and less neutral, leading to a more homogeneous ionization field compared with larger MFP values where the islands are larger with more extreme neutral fractions. It is also clear from these slices that the difference in the ionization field caused by different MFP values is much more significant at higher global ionized fractions and therefore later in reionization; the boxes appear visually nearly identical at $\left< x_{\rm HI} \right> = 0.9$ and show significant differences by $\left< x_{\rm HI} \right> = 0.1$.

\begin{figure}
    \centering
    \includegraphics[width=0.8\linewidth]{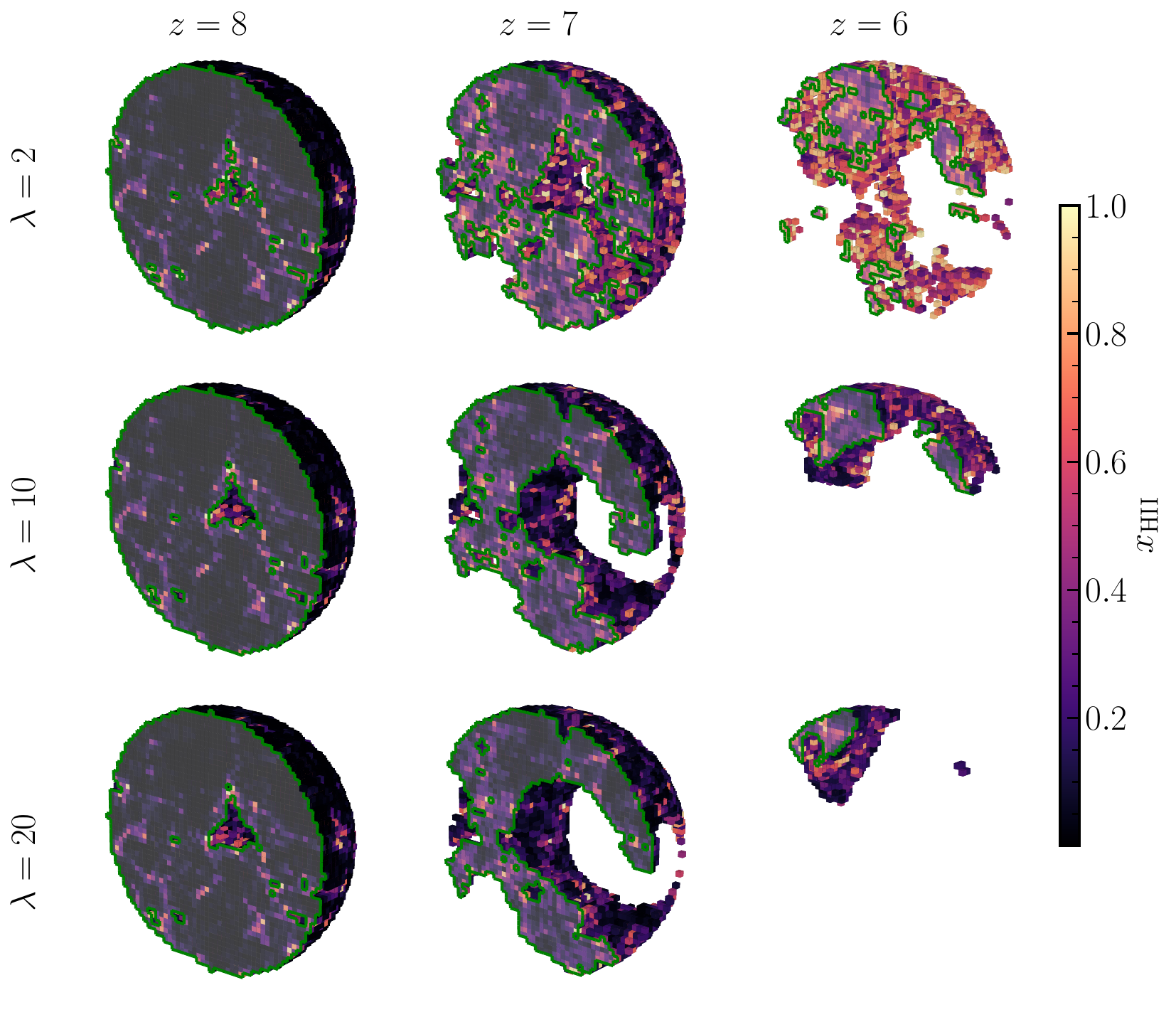}
    \caption{\textbf{A shorter MFP may delay the onset of the percolation threshold during reionization.} Each panel shows a hemisphere from the ionization boxes shown in figure \ref{fig:21cmFASTslices}. Each sphere has a radius of 25 cMpc and corresponds to the middle circle of each panel in figure \ref{fig:21cmFASTslices}, where the slice of each sphere is on the same plane as the slice in figure \ref{fig:21cmFASTslices}. Voxels outlined in green are those on the slice which are not fully ionized (i.e., those visible in figure \ref{fig:21cmFASTslices}). The colorbar shows the ionized fraction of each voxel, where fully ionized voxels are transparent. The same general trends are apparent as in the 2D slices, but the 3D representations reveal more nuance; for example, while neutral regions appear smaller for smaller MFP values in a 2D slice, it is clear from the 3D images that these smaller neutral regions are in fact connected by narrow channels, whereas the neutral regions for larger MFP values seem to be true ``islands.''}
    \label{fig:21cmFASTspheres}
\end{figure}

One common issue with displaying ionized boxes is that 2D representations can be misleading in terms of the distribution of ionized bubbles --- bubbles which appear small in a 2D slice might simply be one end of a much larger bubble which extends either into our out of the plane of the image. To better illustrate the ionization field (specifically on the scales relevant to this work), in figure \ref{fig:21cmFASTspheres} we show hemispheres from the ionization box output by \texttt{21cmFAST}. Each sphere has a radius of 25 cMpc and corresponds to the middle circle plotted in each panel of figure \ref{fig:21cmFASTslices}, where the sphere is sliced on the same plane as that shown in figure \ref{fig:21cmFASTslices}. The voxels located on the slices are outlined in green and fully ionized voxels are transparent so that the neutral regions located \textit{behind} the plane shown in figure \ref{fig:21cmFASTslices} are visible. These examples clearly illustrate the issue of using 2D slices for visualization --- the sphere shown for $\lambda = 2$~Mpc and $z=6$, for example, shows significant structure that is not apparent from the slice alone. The spheres also reveal more nuance in the trend that seems to be the case from figure \ref{fig:21cmFASTslices}; although neutral regions in a given slice tend to be smaller for a smaller MFP, they also appear to be interconnected through narrow neutral channels, whereas leftover neutral regions using a larger MFP seem to appear as true ``islands." Reionization can be thought of as a percolation process, and such interconnectedness is a common feature of reionization models and more generally all percolation processes. As such, \cite{Furlanetto16} find that reionization reaches a ``percolation threshold'' where a single, infinitely large interconnected ionized region appears suddenly. The inverse is also true: the single neutral region which initially comprises the entire Universe is carved out by ionized bubbles until eventually neutral islands separate into individual regions. The study of such a threshold is important to reionization as it helps to provide a precise description of the distribution of ionized and neutral regions. The suggestion of such a state in figure \ref{fig:21cmFASTspheres} indicates that smaller MFPs delay the percolation threshold to larger global ionized fractions and thus to later in reionization. This has interesting implications for studying reionization through the framework of percolation theory in the context of the shorter measured MFP value and thus motivates further study, but a detailed analysis is beyond the scope of the current work. 

\subsection{Mass-dependent ionizing efficiency}\label{ss:mass-depend}

Up to this point we have taken the ionizing efficiency to be a constant. However, this quantity implicitly depends on the average star formation efficiency, which has been demonstrated to scale with halo mass at high redshift (e.g., \citep{trenti_galaxy_2010, tacchella_physical_2013, mason_galaxy_2015, behroozi_simple_2015}). In this section we extend the description of the MFP-dependent ionization criterion described above to include a mass-dependent ionizing efficiency. 

Following the procedure of \cite{furlanetto_minimalist_2017}, we can characterize this relationship with the power law

\begin{equation} \label{eq:mass_dep_zeta}
\zeta = \zeta_0 \left ( \frac{M_h}{M_0} \right )^{\xi},
\end{equation}
where $\xi$ captures the effect of different stellar feedback regulation mechanisms on the star formation history and average star formation efficiency of a halo of mass $M_h$. \cite{sun_2016} found that $\xi \sim $~1/3--2/3 provides a good fit to observations of the luminosity function at $z~>~6$, corresponding to purely momentum- and energy-regulated feedback described in \cite{furlanetto_minimalist_2017}, respectively.

Incorporating this definition of the ionizing efficiency into \ref{eq:MFP-ion-criterion}, we obtain the following ionization criterion for a region of radius $R$ with corresponding mass $M$ (see footnote \ref{fn:mass-radius}):

\begin{equation} \label{eq:smooth_Q}
    1 = \zeta_0 M_0^{-\xi} \frac{\lambda (1 - {\rm e}^{-R/{\lambda}})}{R} \int_{M_{\rm min}}^{M} M_h^{\xi} n(M_h|R, \delta, z) \frac{M_h}{\rho} dM_h
\end{equation}
where $n(M_h | R, \delta, z)$ is the conditional halo mass function, i.e., the probability of finding a halo of mass $M_h$ in a region of radius $R$ and density $\delta$. As shown, the mass-dependent term in the ionizing efficiency must be integrated with the halo mass function. This new expression can be computed numerically and simplifies to \ref{eq:MFP-ion-criterion} in the mass-independent case.

With an extension of this power law, we can also consider how reionization progresses with and without the assumption that star formation rates reach an equilibrium state and therefore have a ``smooth" function across all halos. As discussed in \cite{yamaguchi_extent_2023}, this assumption is implicitly made when we apply a consistent power law across the entire halo mass range. We therefore follow a similar procedure as \cite{yamaguchi_extent_2023} to consider an example of out-of-equilibrium behavior posed by \cite{furlanetto_bursty_2022}, who study the effects of potential time delays between star formation and supernovae feedback. Low-mass halos will experience out-of-equilibrium behavior in the form of ``bursts" of star formation because they are not massive enough to retain the gas ejected by feedback; this causes their star formation to become independent of halo mass. Sufficiently massive halos, which we define as having a mass above the cutoff $M_b$, will not experience this behavior because they can retain their gas. We model this scenario with the ionization criterion
\begin{equation} \label{eq:bursty_Q}
    1 = \zeta_0 \frac{\lambda (1 - {\rm e}^{-R/{\lambda}})}{R} \left (\int_{M_{\rm min}}^{M_b} n(M_h|R, \delta, z) \frac{M_h}{\rho} dM_h  +  \int_{M_b}^M \left( \frac{M_h}{M_b} \right)^{\xi} n(M_h|R, \delta, z) \frac{M_h}{\rho} dM_h \right ),
\end{equation}
where the first term imposes a mass-independent floor for low-mass halos and the second term assumes ``smooth" star formation for more massive halos. We take $M_b$ to be $10^{10} M_{\odot}$ as motivated by \cite{furlanetto_bursty_2022}.

As before, we can invert eq. \ref{eq:smooth_Q} or \ref{eq:bursty_Q} for the density required, as a function of smoothing scale, of a fully ionized region. In this case we must solve for the ionizing barrier numerically. Fortunately, however, this must be done only once for each choice of $\xi$ at each desired redshift (for a given cosmology). Then $\zeta_0$ can be calibrated to achieve the desired global ionized fraction. Figure \ref{fig:mass_dep_zeta_barriers} demonstrates examples of barriers computed using a mass-dependent ionizing efficiency using both the ``smooth'' and ``bursty" star formation models. Qualitatively they maintain a similar shape to the mass-independent case.

\section{Results: effect of the mean free path on the growth of HII regions}\label{s:growth-of-HII}

Armed with our barrier, we can now retrieve the first-crossing distribution and any related quantities. However, as we wish to also compute the one-point PDF of the ionized fraction, further extensions to the model are required, particularly in light of a short MFP. Before describing these modifications in section \ref{s:one-point-PDF-description}, we first explore the implications of the ionization criteria described in \ref{S:semi-ana} for fully ionized regions, for which we require only the first-crossing distribution.

\subsection{First-crossing distribution} \label{s:model}

As with previous photon-counting methods, we rely on the excursion set theory to determine the first-crossing distribution. However, our chosen barrier is a highly non-linear function of scale (see figure \ref{fig:mass_dep_zeta_barriers}) and since no general analytic method of solving for the first-crossing distribution of a barrier of arbitrary shape exists, we use the numerical method outlined in \cite{ZH06} (hereafter \citetalias{ZH06}). We direct the interested reader to \citetalias{ZH06} for a complete derivation and explanation of the method, but we provide a broad overview here.\footnote{For readers interested in reproducing these results, we note here that \citetalias{ZH06} has a typo in their eq. 22, where the prefactor in the third line should be $(1-\Delta_{i,i})^{-1}$ rather than $(1-\Delta_{1,1})^{-1}$ \cite{Zhang_personal}.} 

The solution of \citetalias{ZH06} involves solving for the first-crossing distribution iteratively on a mesh with equal spacing in the variance and assumes that the random walk begins at $(\tilde{S},\tilde{\delta}) = (0,0)$ and ends at some scale $\tilde{S}_{\rm sub}$. Solving our barrier with this method provides us with the first-crossing distribution suitable for describing the size distribution of ionized bubbles, $f(\tilde{S}' |\tilde{S}=0, \tilde{\delta}=0)$, for $0 \leq \tilde{S}' \leq \tilde{S}_{\rm sub}$. However, for the purposes of the one-point PDF (described in section \ref{s:one-point-PDF-description}) we also require the conditional first-crossing distribution, i.e., the first-crossing distribution beginning at some $(S,\delta)$ over a range $S \leq S' \leq S_{\rm sub}$. Solving for the conditional first-crossing distribution is equivalent to performing the method outlined in \citetalias{ZH06} (i.e., solving for the unconditional distribution) but with $\tilde{S}' = S' - S$ and $\tilde{\delta} = \delta - \delta$, where $0 \leq \tilde{S}' \leq S_{\rm{in}}-S$ and the initial density is $\tilde{\delta} = 0$. In terms of the transformed variables, the barrier is thus $\tilde{B}_{\rm{ion}}(\tilde{S}') = B_{\rm{ion}}(\tilde{S}'+S) - \delta$. The random walk is performed over ($N_S$+1) scales $\tilde{S}_{i} = i \Delta \tilde{S}$, where $i = 0,1,...,N_S$ and $\Delta \tilde{S} = (S_{\rm{in}}-S)/N_S$, which, applying the result of \citetalias{ZH06} and shifting back to $S'$, leaves us with values of the first crossing distribution $f(S'_i |S, \delta)$ for each of the discrete values $S_i = i\Delta\tilde{S} + S$. In practice we use $N_S = 2000$, which takes under $2\,$s to run. For ease of notation we drop the subscript $i$ and refer to the conditional first crossing distribution without reference to the discrete values: $f(S' |S, \delta)$.

\subsection{Size distribution of HII regions}

The first-crossing distribution allows us to describe the size distribution of ionized bubbles. As in \citetalias{FZH04}, we compute $\bar{Q}^{-1} V' dn/d \ln R'$, where 

\begin{equation}
    \frac{dn}{dM_{\rm i}'} = \frac{dn}{dR'} \frac{dR'}{dM_{\rm i}'} =\frac{\bar{\rho}}{M_{\rm i}'} f(S' | 0,0) \left| \frac{dS'}{dM_{\rm i}'} \right|
\end{equation}
is the mass function of ionized bubbles \citep{Sheth98}, $V'$ is the comoving volume of a bubble of mass $M_{\rm i}'$ (corresponding to radius $R'$; see footnote \ref{fn:mass-radius}), and the expression is normalized by the volume-filling fraction of ionized bubbles:

\begin{equation} \label{eq:simple_Q}
    \bar{Q} = \int dM_{\rm i}' \frac{dn}{dM_{\rm i}'} V'.
\end{equation}

\begin{figure}
    \centering
    \includegraphics[width=1.0\linewidth]{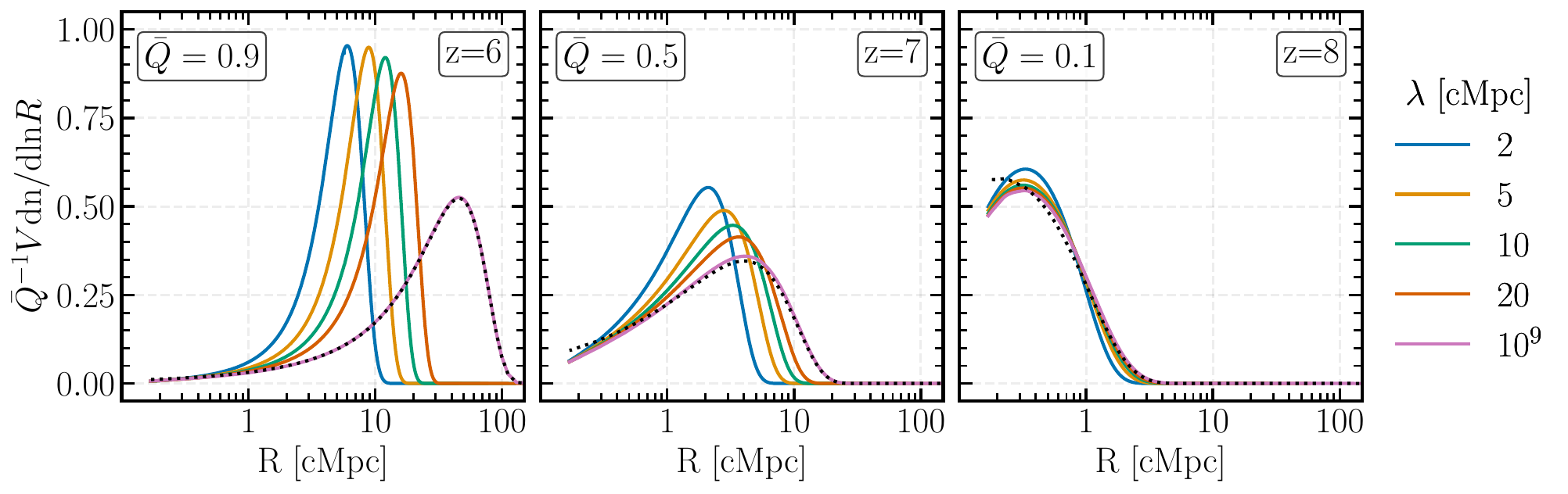}
    \caption{\textbf{The sizes of ionized bubbles are much more sensitive to the MFP during the later stages of reionization.} Each panel shows the size distributions $\bar{Q}^{-1} V dn/d\ln R$ of fully ionized bubbles at different redshifts and volume-filling fractions $\bar{Q}$ at various choices for the MFP (colors). All cases use $\xi = 0$. Also shown are the standard photon-counting results using the method outlined in \citetalias{FZH04} (black dotted lines). As $\lambda~\rightarrow~\infty$, the result of our model approaches that of \citetalias{FZH04}, as expected. Slight deviations occur because the solution of \citetalias{FZH04} approximates the barrier as linear, whereas our solution does not. As the global ionized fraction increases towards the end of reionization, the differences in the size distributions for different MFP lengths become much more pronounced, indicating that the MFP has an increasing effect on the topology as reionization progresses.}
    \label{fig:size-dist}
\end{figure}

In figure \ref{fig:size-dist} we show the size distribution of ionized bubbles at several volume-filling fractions and MFP lengths. For comparison, we also show results using the traditional barrier which does not consider the MFP. The effect of the MFP on the size distribution of ionized bubbles is apparent; towards the end of reionization, at $\bar{Q} = 0.9$ in our example, shorter MFP lengths result in substantially smaller ionized bubbles. Using MFP values within the range found by \citetalias{B21}, we find that the mean bubble size can be up to around an order of magnitude smaller compared with that found using the traditional photon-counting criterion (which ignores the MFP). Compared with methods which apply a maximum radius in the excursion set, such as \cite{Furlanetto05}, which tend to find a bubble size distribution that peaks dramatically at the maximum radius, this method results in a smoother distribution and allows for ionized bubbles larger than the MFP length. We find that the sizes at the end stages of reionization are also quite sensitive to the particular value of the mean free path, even within the relatively narrow range of values measured in \citetalias{B21}, while as the volume-filling fraction decreases, the sizes become less sensitive to the MFP. This is to be expected; when bubble sizes are much smaller than the mean free path scale, the precise value of the mean free path has little effect (see section \ref{S:semi-ana} for a discussion on the subtleties of the MFP as defined in the ionization criterion). However, once ionized bubbles reach the mean free path scale the ability for larger bubbles to form is severely restricted. 

Although it is difficult to define the precise extent of ionized bubbles in 3D realizations (because reionization is a percolation process; see e.g., \citep{Lin2016, Furlanetto16, Giri21} and figure \ref{fig:21cmFASTspheres}), bubbles are found to be larger in semi-numerical simulations when compared with excursion-set--based analytic models by a factor of a few \cite{Mesinger07, Lin2016}, where in general better agreement is found towards the later stages of reionization. We expect our results to be no different in this regard, though the qualitative conclusions in this section should remain unchanged.

\subsection{Fraction of HII regions and valid scales of the model}
\begin{figure}
    \centering
    \includegraphics[width=0.66\linewidth]{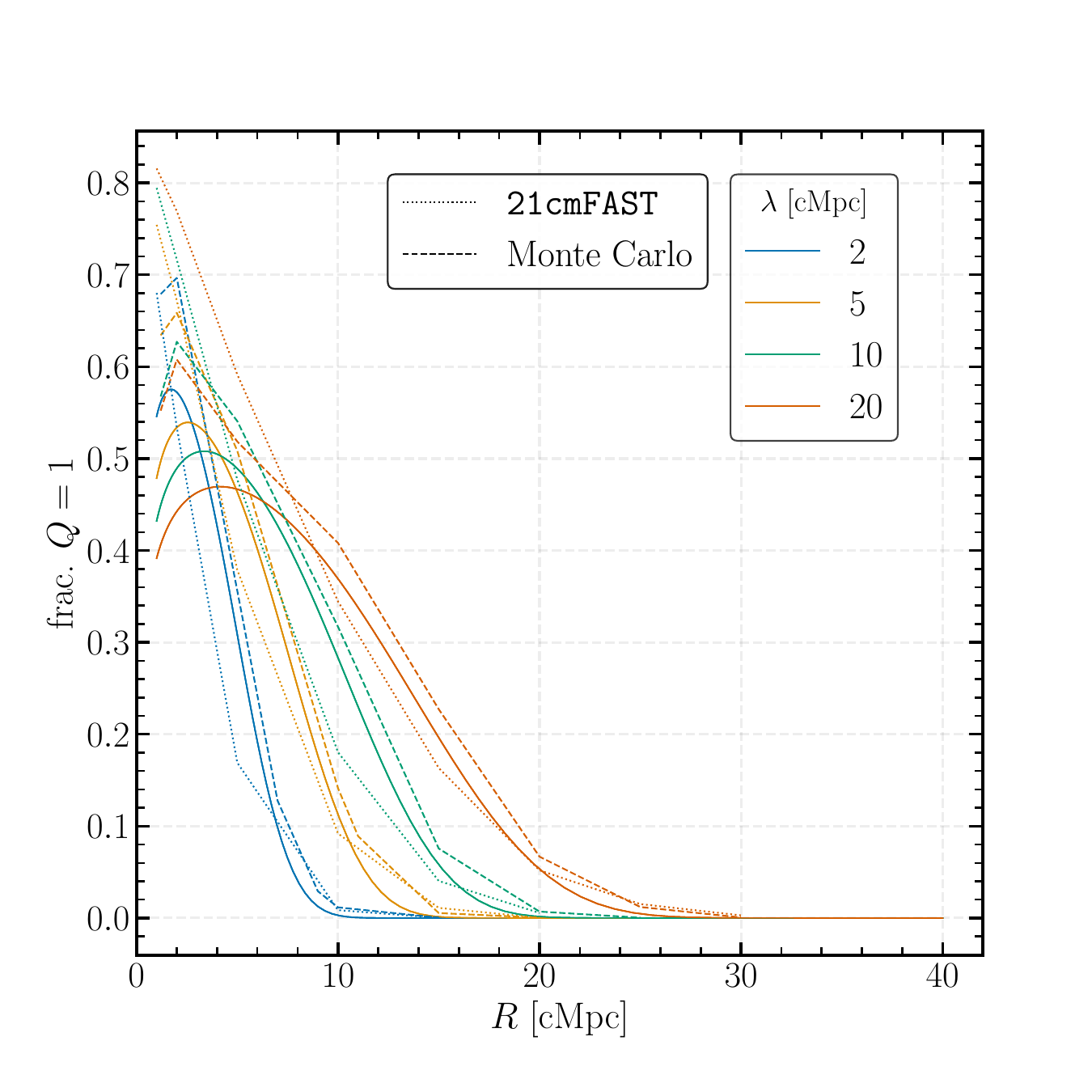}
    \caption{Fraction of fully ionized regions as a function of scale. Results are shown for a variety of MFP values (colors). All cases are at $z=6$ and $\left< x_{\rm HI} \right> = 0.1$, and a constant ionizing efficiency ($\xi = 0$). Results are shown using our model (solid lines), $\texttt{21cmFASTv1}$ with the same ionization criterion (dotted lines), as well as a Monte Carlo version of our model, which accounts for a finite number of subregions within each larger region (dashed lines; see section \ref{s:one-point-PDF-description} for more details on the use of subregions and appendix \ref{a:21cmFAST-differences} for details on the Monte Carlo version of the model). While this fraction should be monotonic, our model (both the analytic and Monte Carlo version) produces a turnover at small radii. This is because of the reduced amount of scatter in the mapping from density to ionized fraction on the smoothing scale (see appendix \ref{a:21cmFAST-differences} for more detail). In the analytic model there is no scatter, and the turnover is most severe. The Monte Carlo model includes some sources of scatter, resulting in a less severe turnover at smaller radii. In all cases, good agreement is shown at large scales ($R \gtrsim 15$ cMpc), where the number of subregions grows large and the scatter in the ionized fractions of the regions as a function of their density grows small, which are the scales relevant to this work.}
    \label{fig:PQ1_funcR}
\end{figure}

Related to but distinct from the size distribution of ionized bubbles, we also explore the fraction of regions as a function of scale which are fully ionized. For our model, this is simply the fraction of regions on a given scale which have a density larger than the barrier at that same scale. In figure \ref{fig:PQ1_funcR} we show the probability for a region to be fully ionized region as a function of its radius. We show results for a variety of MFP values. In all cases $z=6$ and $\zeta_0$ has been calibrated so that the global neutral fraction is $\left< x_{\rm HI} \right> = 0.1$. The probability at an infinitely large radii is 0 (as the global ionized fraction is not 1) and increases as radius decreases. The short MFP has a significant effect on the fraction of regions of a given radius which are fully ionized. For example, at a radius of $R = 20$ cMpc, an MFP of $\lambda = 20$ cMpc results in $\sim 5$\% of regions being fully ionized, while with an MFP at the mean value found by \citetalias{B21} (5~cMpc), such regions are vanishingly rare. This trend simply reflects the increased bubble size for larger MFP values at a fixed global neutral fraction. 

The probability of finding a fully ionized region should increase monotonically with decreasing radius, converging to the global ionized fraction (0.90 in the case of figure \ref{fig:PQ1_funcR}) at sufficiently small radii. Clearly this is not the case in our model. This is most likely due to approximations in our model which remove scatter which we expect to be present, and which is present in \texttt{21cmFAST}. This scatter can behave in subtle ways to alter the ionized fractions of smoothed regions. We describe some important differences between the two models in appendix \ref{a:21cmFAST-differences}, which we find to be most significant at small radii. Given these discrepancies, we focus this work only on scales $R\gtrsim 15$ cMpc where the two methods show good agreement, and which are also the scales most relevant to upcoming observations of the 21-cm signal.

\section{Modeling the one-point PDF} \label{s:one-point-PDF-description}

The size distribution of ionized bubbles obtained from the first-crossing distribution is a useful statistic as it describes the very features which determine the topology of the ionization field --- the bubbles themselves. However, in practice, measurements of the size of ionized bubbles are difficult. As mentioned above, even within simulations there is ambiguity in the way in which individual bubbles are distinguished from one another owing to the complex shapes and interconnectedness of ionized regions. For this reason, and particularly for the sake of comparisons with observations, for the remainder of this work we choose to focus our model on describing partially ionized regions of a fixed size, which will allow us to predict one-point statistics of the ionization field and 21-cm signal. 

To do this, as described in the following sections, we will continue to apply the excursion-set approach. Briefly, we recreate the approach of \texttt{21cmFAST} analytically. That is, we imagine dividing a region of a given radius $R$ into constituent smaller `subregions' (akin to voxels in \texttt{21cmFAST}), each of which is sufficiently small compared with the MFP length. We then build the range of possible outcomes for the ionized fraction of these subregions conditionally based on the density of the larger region, where their expected mean gives the average ionized fraction of the larger region. Using the known probability distribution of the underlying density field at radius $R$ then allows us to reconstruct the PDF of the ionization field. We step through this process in detail in the following sections.

\subsection{Use of subregions}\label{ss:subregions}

We wish to find the ionized fraction of some region of interest with a density $\delta$ and radius $R$ (or scale $S$). With an eye towards direct observational comparison, one could imagine that such a region represents a patch of the sky (with some thickness), over which the ionized fraction (or, more likely, the 21-cm signal strength) is averaged, or \textit{smoothed}. We will refer to such regions as smoothed regions and the radius $R$ (or $S$) as the smoothing radius (or scale).

We imagine dividing the region into smaller constituent `subregions' at a scale $S_{\rm sub} > S$ (or equivalently $R_{\rm sub} < R$), such that the average of the ionized fractions of these subregions gives the ionized fraction of the region of interest. The size of these subregions should be chosen to be sufficiently small compared with the MFP such that homogeneity of photons is a good approximation at these scales. We then imagine performing a separate random walk on each of these subregions, ionizing each if its trajectory crosses the barrier at any $S' < S_{\rm sub}$. If a trajectory does not cross the barrier at any point prior to reaching the subregion scale, its ionized fraction is set to some fraction using

\begin{equation} \label{eq:MFP-Q}
    Q_{\rm{sub}}(\delta') = \zeta f_{\rm{coll}}(\delta' | S_{\rm sub}),
\end{equation}
where all quantities are those of the subregion. Note that for a mass-dependent ionizing efficiency, $f_{\rm coll}$ differs from the standard definition, and is instead defined through eqs. \ref{eq:smooth_Q} and \ref{eq:bursty_Q}. In such cases we also include the prefactor which accounts for the MFP that is present in these two equations. When the radius $R$ is small compared to $\lambda$ this prefactor is near unity and so in practice we find that whether or not it is included does not substantially change our results. This method is equivalent to per-pixel random walks in \texttt{21cmFAST} or other similar seminumerical methods, using the method of ionizing only the central cell.

In general all random walks begin at an infinite radius and at the average density of the Universe, where the variance and overdensity are both 0. In practice, semi-numerical models must have a finite starting scale, which can affect various statistics, depending on the physical size of the simulation box \cite{Barkana04}. One benefit of an analytic excursion set approach is that we are free to choose this scale arbitrarily by solving for the conditional first-crossing distribution, allowing us to validate the simulation approach. We quantify the effect of finite simulation boxes in appendix \ref{s:finite_box}, but generally find that boxes with side lengths $\gtrsim 300$ cMpc are sufficient to describe the one-point statistics at radii $\lesssim 50$ cMpc, at least as far as it concerns the excursion set. Any random walk which describes a subregion contained within our region of interest must also pass through the density of the region $\delta$ on the scale of the region $S$. If we define a quantity $P_{\rm cross}(\delta|S)$ as the probability that a random walk for a subregion crosses the barrier at some scale $S' < S_{\rm sub}$ \textit{given} that it crosses through the point $(S,\delta)$, then the ionized fraction of our region of interest is

\begin{equation} \label{eq:Q_of_d}
    Q(\delta|S) = P_{\rm cross}(\delta|S) + [1 - P_{\rm cross}(\delta|S)] \mathbb{E}_{Q, \rm nc}(\delta|S),
\end{equation}
where $\mathbb{E}_{Q, \rm nc}(\delta|S)$ is the expected value of the ionized fraction of all  subregions whose associated random walks do \textit{not cross} the barrier at $S' < S_{\rm sub}$. We now describe the calculation of each term in eq. \ref{eq:Q_of_d}.

\paragraph{$\mathbf{P_{\rm cross}:}$} We can break down $P_{\rm cross}(\delta|S)$ into two parts: given that a trajectory passes through $(S,\delta)$, we have (\textit{i}) the probability that it crosses the barrier at $S' < S$, $P_{\rm cross, S' < S}(\delta|S)$, and (\textit{ii}) the probability that it crosses at $S < S' < S_{\rm sub}$, $P_{\rm cross, S' > S}(\delta|S)$. As the random walks of the excursion set method are uncorrelated, the two probabilities are independent. $P_{\rm cross}(\delta|S)$ is the probability that a random walk crosses at one of the smaller \textit{or} larger scales (inclusively), and so:

\begin{equation} \label{eq:Pcross}
    P_{\rm cross}(\delta|S) = P_{\rm cross, S' > S}(\delta|S)[1 - P_{\rm cross, S' < S}(\delta|S)] + P_{\rm cross, S' < S}(\delta|S).
\end{equation}

$f(S'|\delta, S)$ is the conditional first-crossing distribution which describes the fraction of trajectories which start at a point $(S,\delta)$ and first cross the barrier at a scale $S'$ (see section \ref{s:model}). We have that

\begin{equation} \label{eq:cross_largeS}
    P_{\rm cross, S' > S}(\delta|S) = \int_S^{S_{\rm sub}} f(S'|\delta, S) dS',
\end{equation}
and, using the general product rule and assuming we start our random walks from an infinite radius where $(\delta_0,S_0) = (0,0)$,
\begin{equation} \label{eq:cross_smallS}
    P_{\rm cross, S' < S}(\delta|S) = P_0(\delta-0,S-0)^{-1} \int_0^{S} f(S'|0,0) P_0(\delta - B(S'),S-S') dS',
\end{equation}
where

\begin{equation} \label{eq:transition-prob}
    P_0(\Delta \delta, \Delta S) = \frac{1}{\sqrt{2 \pi \Delta S}} \exp\left[ \frac{-\Delta \delta^2}{2\Delta S} \right]
\end{equation}
describes the probability to find a region of density $\delta_{\rm in}$ on a scale $S_{\rm in}$ within a region of density $\delta_{\rm out} = \delta_{\rm in} - \Delta \delta$ on a scale $S_{\rm out} = S_{\rm in} - \Delta S$, a fundamental probability in the excursion set formalism. $P_0(\delta - B(S'),S-S')$ is thus the probability to transition from the barrier at $S'$, ($S',B(S'))$ to the point of interest $(S,\delta)$ and $P_0(\delta - 0,S-0)$ is the probability to pass through $(S,\delta)$ starting from $(0,0)$. The integral term gives the probability to cross the barrier \textit{and} pass through the point of interest, while the first term gives the probability to pass through the point of interest regardless of whether the barrier is crossed. The quotient (through the product rule) gives the probability that a trajectory crosses the barrier at $S' < S$ \textit{given} that it passes through $(S,\delta)$. 

\paragraph{$\mathbf{\mathbb{E}_{Q, \rm nc}(\delta|S)}$:} To compute $\mathbb{E}_{Q, \rm nc}(\delta|S)$, we need the probability distribution of densities at $S_{\rm sub}$ for all trajectories which pass through $(S,\delta)$ but which do not cross the barrier at $S' < S_{\rm sub}$. The expected ionized fraction then follows, as the ionized fraction of a subregion in our model is a deterministic function of its density (eq. \ref{eq:MFP-Q}).

First, consider only scales $S' > S$. For trajectories beginning at $(S,\delta)$, the probability of passing through $(S_{\rm sub},\delta')$ without having crossed the barrier at any intermediate scale is 

\begin{equation}
    \tilde{P}(\delta'|S_{\rm sub}) = P_0(\delta'-\delta,S_{\rm sub}-S) - \int_S^{S_{\rm sub}} f(S';\delta,S) P_0(\delta' - B(S'),S_{\rm sub}-S') dS'.
\end{equation}
However, we must also consider the trajectories at larger radii with scales $S' < S$. Again, we must include a factor which accounts for trajectories which cross at $S' < S$, given that they cross through $(S,\delta)$, which is precisely Eq. \ref{eq:cross_smallS}. So, the probability to find an inner region of density $\delta'$ within a region of $(S,\delta)$ given that it does not cross the barrier at any $S' < S_{\rm sub}$ is

\begin{equation}
    P(\delta'|S_{\rm sub}) = \frac{1 - P_{\rm cross, S' < S}(S,\delta)}{1-P_{\rm cross}(S,\delta)} \tilde{P}(\delta'|S_{\rm sub}).
\end{equation}
A more complex treatment at $S'<S$ is not needed as we know that all trajectories pass through $(S,\delta)$ and because the random walks are uncorrelated. Thus, the only information we need is the fraction of trajectories which reach the point of interest without having been ionized at a smaller $S$. The expected value is then
\begin{equation}
    \mathbb{E}_{Q,\rm nc}(\delta|S) = \int_{- \infty}^{\infty} Q_{\rm sub}(\delta') P(\delta'|S_{\rm sub}) d \delta'.
\end{equation}
We note that our treatment here results in a deterministic relationship between a region's density and its ionized fraction. In reality, there will be some scatter in this relationship. We discuss this in more detail in appendix \ref{a:21cmFAST-differences}.

\subsection{Computing the one-point PDF}

Armed with a mapping from $(S,\delta)$ to ionized fraction, the probability of finding a region on scale $S$ with ionized fraction $Q$ then simply follows from the probability of finding a region of a given density:
\begin{equation}\label{eq:PofQ}
    P_Q(Q|S) = P_0(\delta, S) \frac{d \delta}{d Q(\delta|S)},
\end{equation}
where $P_0(\delta,S)$ is given by eq. \ref{eq:transition-prob}.

In practice, for a given smoothing scale, we compute $Q(\delta|S)$ for a representative sample of densities and interpolate between them using a cubic spline. $Q(\delta|S)$ tends to be sigmoidal except for a sharp transition at the barrier where the first derivative is not continuous. Such a transition makes smooth interpolation such as that with a cubic spline inaccurate at the very high end, where the interpolated function tends to underestimate the ionized fraction. As such, we interpolate only up to some density close to the barrier, meaning that we do not recover the \emph{shape} of the distributions at extremely high ionized fractions (generally $>99\%$ ionized), though we can still determine the total fraction of regions that are $>99\%$ ionized and fully ionized. This does not affect our results otherwise.

\subsection{Comparison to previous analytic models}

Previous analytic models have explored computing partially-ionized regions for the purposes of describing the one- (or higher-) point PDF \citep{FZH04, Barkana08}. However, in these cases only radii \textit{larger} than the smoothing scale are considered. That is, a random walk is performed down to the smoothing scale and then, if the region is not fully ionized upon reaching this scale, its ionized fraction is set using $Q = \zeta_0 f_{\rm coll}$ (or a similar expression). As such, the sub-structure of regions on the smoothing scale are not considered, and instead partial ionizations are set only using the averaged information on the smoothing scale. This approximation assumes that all photons are homogeneously spread throughout the region. Given the short MFP value, this approximation is no longer valid. By dividing regions into subregions on a scale which is small compared with the MFP length, our method is able to better account for an inhomogeneous distribution of density (and thus sources and photons), albeit only in a statistical sense. \cite{Trapp23} employ a similar method of subregions in the context of constraining reionization through observations of Lyman-$\alpha$ emitters. However, for the sake of simplicity, in this case the authors ignore any other scales entirely, foregoing a random walk method and considering only the density on the subregion scale. 

\section{Results: the one-point PDF of the ionization field} \label{s:one-point-PDF}

We now turn our attention more strictly to the one-point PDF and its dependence on the various parameters in the model.

\subsection{Mean free path} \label{ss:MFP}
\begin{figure}
    \centering
    \includegraphics[width=1.0\linewidth]{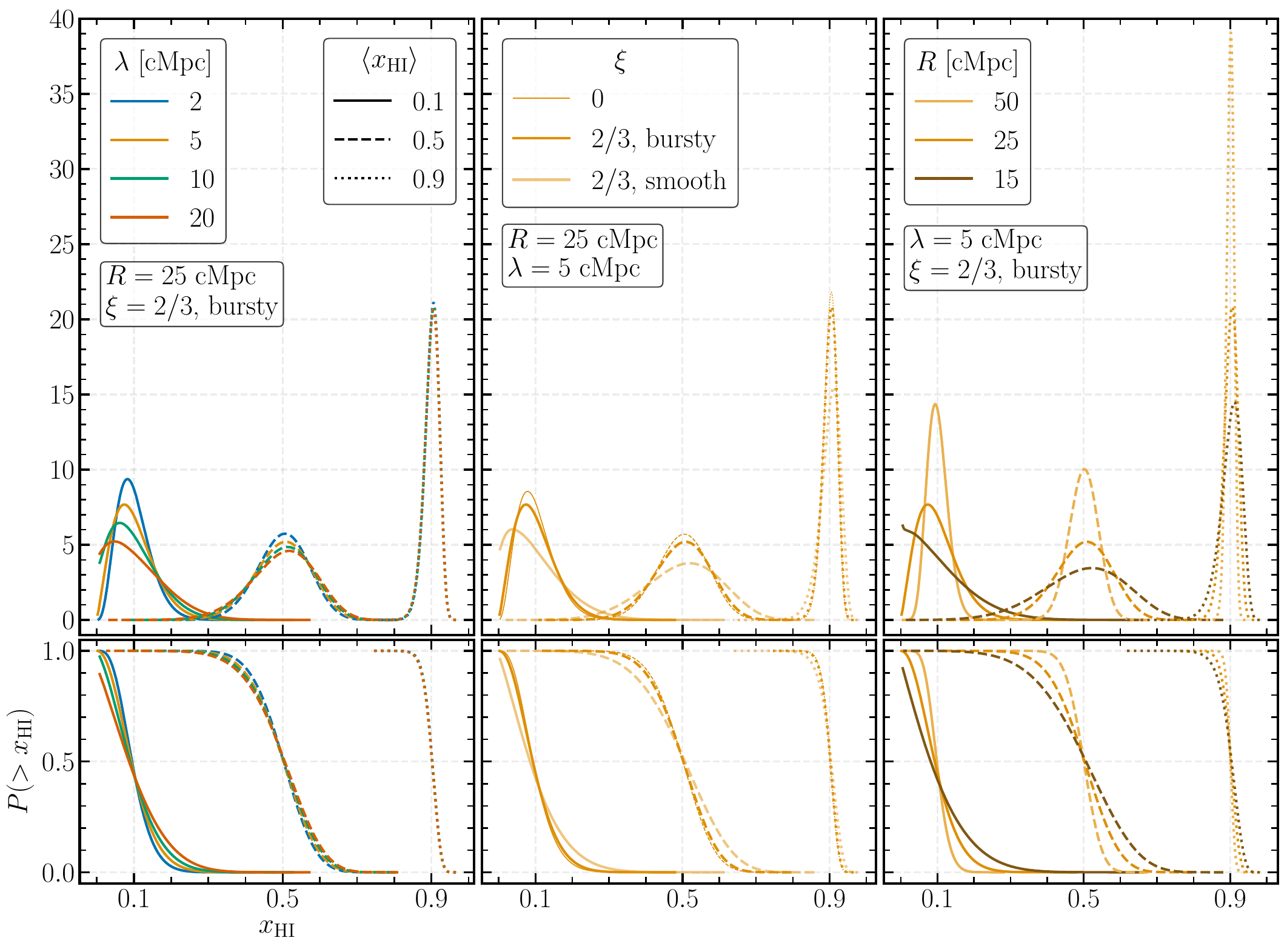}
    \caption{\textbf{The one-point PDF is much more sensitive to the MFP towards the end stages of reionization, and is also sensitive to the power-law index of the mass-dependence of the ionizing efficiency and the smoothing scale across redshifts.} Top panels show one-point PDFs of the ionization field and bottom panels show the corresponding CDFs at several MFP lengths (colors; left panels), cases for $\xi$ (line thicknesses; middle panels), and smoothing scales $R$ (color shades; right panels). In each pair of panels, for groups of distributions from left to right (also shown in different line styles), the global neutral fraction is fixed to $\left< x_{\rm HI} \right> = 0.1$, 0.5, and 0.9, respectively, by calibrating $\zeta_0$. Note that the CDFs do not all have a maximum value of 1 because fully ionized regions are not included in PDFs.}
    \label{fig:PDF-all}
\end{figure}

In the left panels of figure \ref{fig:PDF-all} we show the one-point PDF for several choices of the MFP: those representing the mean and error bars of the value reported in \citetalias{B21} at $z=6$, as well as a larger value of $\lambda = 20$ cMpc. We show the result for several global ionized fractions. Because the model is much more sensitive to the global ionized fraction rather than the particular choice of redshift, rather than allowing the natural redshift evolution of our model, we instead calibrate $\zeta_0$ at each redshift independently so that the global neutral fraction is fixed for all choices of $\lambda$. This plot highlights the increased sensitivity of the one-point PDF to the MFP towards the end stages of reionization, at lower mean neutral fractions; when the Universe is only 10\% ionized the MFP values used here produce visually indistinguishable one-point PDFs, while the widths of the distributions become more distinct as reionization progresses, with larger MFP values producing wider PDFs. In the bottom panel we show the corresponding cumulative distribution function (CDF) of each PDF, which highlights the same trend. The reionization topology is determined by the balance between $\zeta_0$ and $\lambda$; for a fixed global ionized fraction, larger MFP values will have a smaller value of $\zeta_0$. This will lead to fewer, but larger ionized bubbles, with more extreme neutral regions leftover. This is in contrast to smaller MFP values where we expect more numerous but smaller ionized bubbles, and leftover regions which have less extreme neutral fractions. The latter case results in a more homogeneous ionization field, and so produces a narrower one-point PDF. This trend is also reflected in the simulation boxes shown in figures \ref{fig:21cmFASTslices} and \ref{fig:21cmFASTspheres}: larger MFP values lead to wider distributions, since randomly chosen spheres are more likely to lie either within more ionized or more neutral regions, whereas smaller MFPs lead to an overall more homogeneous distribution (for a fixed smoothing scale) and therefore a narrower PDF. At either low or high neutral fractions, an increase in width is accompanied by a more pronounced skewness of the distribution.

\subsection{Mass-dependence of $\zeta$} \label{sss:zeta}
The middle panels of figure \ref{fig:PDF-all} show the same distributions but for different values of $\xi$, the mass-dependence of the ionizing efficiency $\zeta$. In this case, all curves are for an MFP of approximately the mean value measured in \citetalias{B21} ($\lambda$ = 5 cMpc). Across all redshifts shown, the power law mass-dependence of the ionizing efficiency results in wider distributions, and using the smooth star formation model has an even more significant effect. This makes sense, as the smooth star formation model more significantly weights higher mass halos compared with the other two methods, as is reflected in the barriers in figure \ref{fig:mass_dep_zeta_barriers}. When high-mass halos are weighted more, we expect that ionized bubbles should be fewer in number but larger (for a fixed global neutral fraction). Using the same argument as section \ref{ss:MFP}, this results in a wider one-point PDF.

\subsection{Smoothing scale}\label{sss:smoothing_scale}

In the right panels of figure \ref{fig:PDF-all} we explore the effect of the smoothing radius $R$ on the one-point PDF. In order to contextualize our results, we choose smoothing scales which are relevant to upcoming direct images of the Epoch of Reionization from telescopes such as HERA and SKA. Unsurprisingly, at a fixed global ionized fraction, smaller smoothing radii result in wider one-point PDFs. At a fixed radius, the width of the distribution changes with the global ionized fraction as the topology and ionized bubble size distribution evolve: at higher $\left< x_{\rm HI} \right>$, a fixed smoothing radius is larger compared to the typical bubble size, and so results in a narrower distribution. As the bubble sizes grow and the typical bubble size approaches the smoothing radius, the distributions become wider. This has well-known implications for observational prospects; smaller resolutions result in a signal with higher contrast, but also typically increase the noise in radio interferometers. The optimal choice of resolution will thus depend on the balance between these two factors. We return to a more detailed consideration of sensitivity and smoothing scale in section \ref{s:sensitivity}.

\subsection{Redshift evolution}

\begin{figure}
    \centering
    \includegraphics[width=0.66\linewidth]{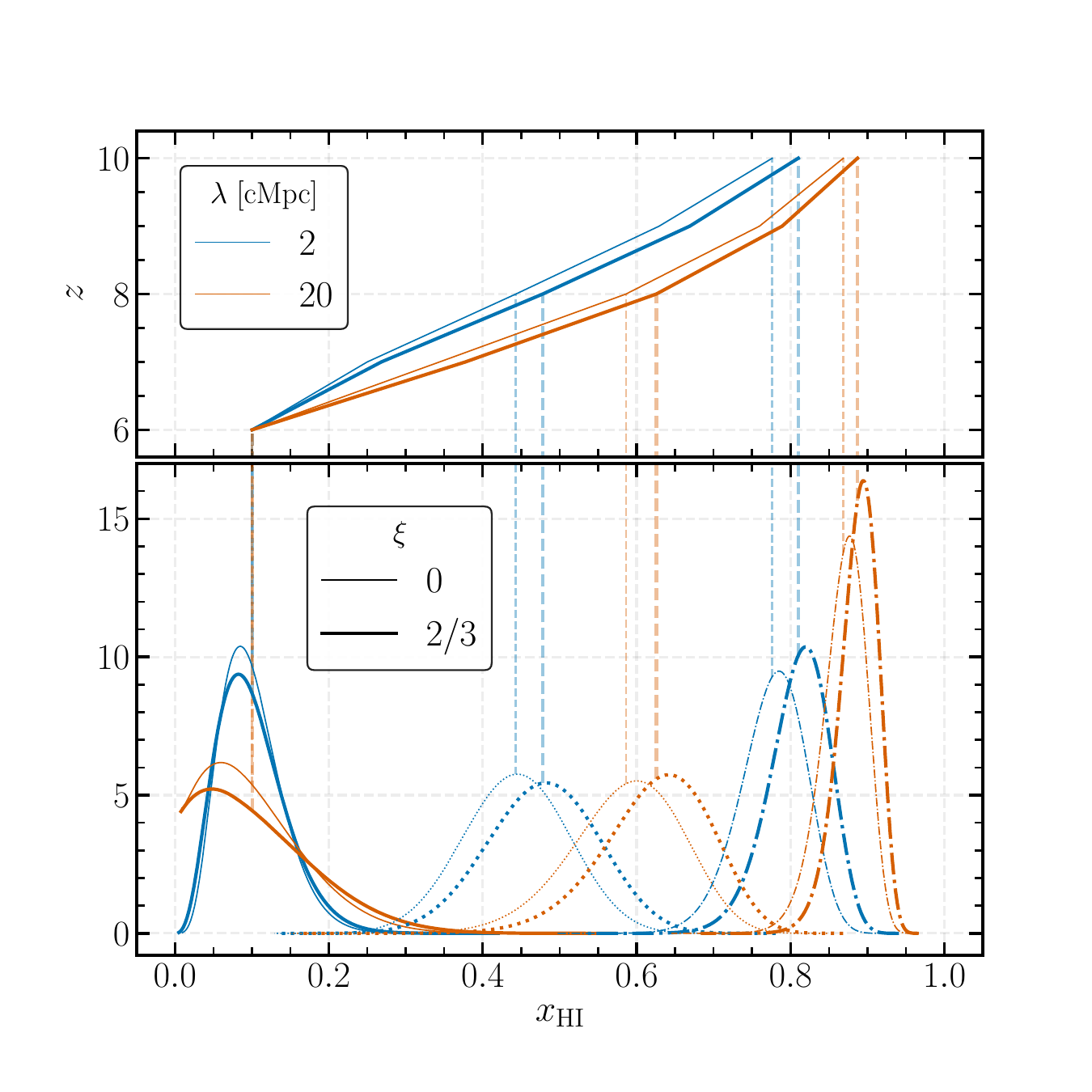}
    \caption{\textbf{The MFP and mass-dependence of the ionization efficiency affect the global ionization history.} The bottom panel shows one-point PDFs from left to right for $z=$ 6 (solid), 8 (dotted), and 10 (dot-dashed) for $\lambda=$ 2 cMpc (blue) and 20 cMpc (orange). Results are shown for $\xi=$ 0 (thin lines), and $2/3$ using the bursty model (thick lines). $\zeta_0$ is calibrated so that the global neutral fraction is $\left< x_{\rm HI} \right> = 0.1$ in all cases at $z=6$ and then fixed at this value for the higher redshifts, corresponding to a redshift-independent ionizing efficiency. The top panel shows the redshift as a function of the global ionized fraction. Vertical lines indicate the global neutral fraction of each one-point PDF shown. Larger MFP values and a larger mass-dependence of the ionization efficiency both result in a more rapid change in the ionization history, though an order of magnitude change in the MFP in this case results in a maximum difference of $\left< Q \right>$ of only $\lesssim 0.2$ around $z=8$. All cases are for a smoothing radius of $R = 25$ cMpc.}
    \label{fig:global-history}
\end{figure}

Although throughout this work we generally calibrate our results to a given global ionized fraction (rather than a given redshift), it is also interesting to explore the effect of various parameters on the natural redshift evolution of the model. For simplicity we assume that $\zeta$ is independent of redshift and calibrate $\zeta_0$ so that $\left< x_{\rm HI} \right> = 0.1$ at $z=6$ and then fix $\zeta_0$ to this value for higher redshifts. The resulting one-point PDFs are shown in the bottom panel of figure \ref{fig:global-history}. Vertical lines show the mean neutral fraction for each distribution, with matching color and linestyle. The top panel shows the redshift as a function of neutral fraction. We find that longer MFP values result in a more rapid reionization (on average across the redshift ranges shown), though an order of magnitude difference in the MFP (2 vs. 20 cMpc) results in a maximum difference of $\lesssim 0.2$ in the global ionized fraction around $z=8$. Interestingly, the case with a longer MFP value initially proceeds \textit{less} rapidly until around the midpoint of reionization where it becomes more rapid compared with the case with a smaller MFP value. This is likely due to the balance between $\lambda$ and $\zeta_0$ --- in order to achieve the same global neutral fraction at $z=6$, the case with a longer MFP must have a smaller value for $\zeta_0$. At high redshifts where the sensitivity to the MFP length is minimized, this results in a reduced growth rate of ionized bubbles. Once the typical bubble size reaches the MFP length, however, the growth of ionized bubbles is restricted. This limit is reached earlier in scenarios with a smaller value for $\lambda$, for which reionization thus proceeds more slowly in the later stages. A mass-dependent ionizing efficiency also results in a slightly more rapid reionization compared with a mass-independent $\zeta$ (with the bursty model), reflecting the trend that higher mass halos evolve more rapidly. However, given that the MFP in our model does not evolve with time, the details of the redshift evolution presented here should be interpreted with caution (see section \ref{S:semi-ana} for some discussion on the subtleties of a redshift-independent MFP parameter).

\subsection{The one-point cumulative distribution function}
\begin{figure}
    \centering
    \includegraphics[width=1.0\linewidth]{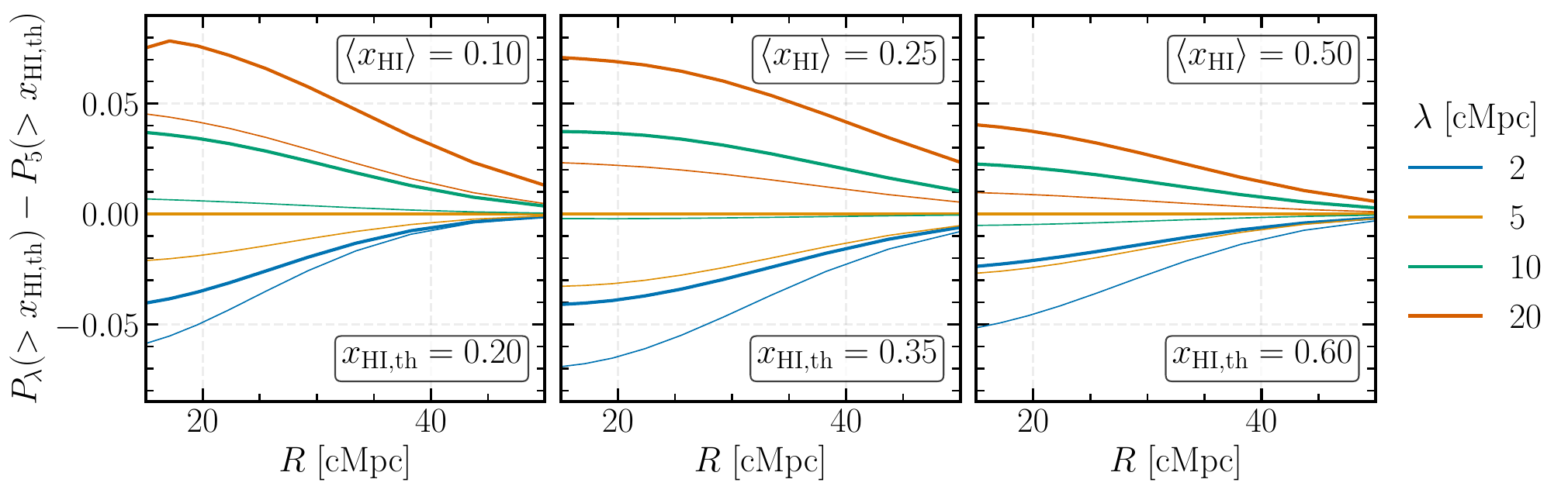}
    \caption{\textbf{The cumulative distribution function of the one-point PDF may offer a promising observational method of distinguishing between reionization models including the MFP length.} Curves show the difference between the CDFs for various MFP and mass-dependences and that for $\lambda=5$~cMpc and $\xi=2/3$ (bursty) evaluated at a single threshold neutral fraction $x_{\rm HI,th}$. Colors show different values of the MFP, and line widths indicate $\xi=0$ (thin lines) and $\xi=2/3$, bursty (thick lines). Different panels show results for global neutral fractions of $\left< x_{\rm HI} \right> = 0.10$, 0.25, and 0.50. In each case the threshold neutral fraction is $x_{\rm HI,th} = \left< x_{\rm HI} \right> + 0.1$, which is approximately +1--2$\sigma$ from the mean in each case, though the exact value depends on global ionized fraction, smoothing scale, and MFP. All models show differences of $\sim$ a few tenths in the CDF at smoothing scales achievable by current and upcoming radio interferometers (see section \ref{s:sensitivity}), indicating that the CDF may offer a promising observational method for distinguishing between reionization models including the MFP length.}
    \label{fig:CDF_funcR}
\end{figure}

To further illustrate the sensitivity of the one-point PDF to the various parameters in the model, in figure \ref{fig:CDF_funcR} we show the differences in the one-point CDF evaluated at a particular choice of ``threshold'' neutral fraction $x_{\rm HI,th}$ (i.e., the fraction of smoothed regions with a neutral fraction larger than $x_{\rm HI,th}$) as a function of the smoothing scale and at various values of $\lambda$ and cases for $\zeta$. We show results at several global neutral fractions $\left< x_{\rm HI} \right>$. In each case the threshold neutral fraction is $x_{\rm HI,th} = \left< x_{\rm HI} \right> + 0.1$, which is approximately $+1$--$2 \sigma$ from the mean, though it varies depending on the global neutral fraction, smoothing scale, and MFP value. In these cases the CDF shows a difference of $\sim$ a few tenths between both different MFP values and mass-dependences at smoothing scales relevant to current and upcoming radio interferometers (see section \ref{s:sensitivity}), indicating that one-point statistics may be useful in distinguishing the MFP length. For example, the left panel shows that at $R=25$~cMpc, the CDF evaluated at $x_{\rm HI,th} = 0.20$ has a value that is $\sim 0.06$ larger for an MFP of 20 cMpc than for an MFP of 5 cMpc. We note that the precise choice of $x_{\rm HI, th}$ (compared with the mean value) will affect how distinct the CDFs for each choice of parameters are. This choice could in principle be optimized, though that was not done in this case as figure \ref{fig:CDF_funcR} simply serves as a representative example.

\section{Prospects for 21-cm tomography} \label{s:sensitivity}

While many existing efforts are focused on the global signal or power spectrum, tomography remains the ultimate goal of 21-cm observations, offering the greatest prospects for understanding the Epoch of Reionization. Beyond any observational effects (including the foreground wedge for 21-cm observations in particular; see e.g., \cite{Liu14a, Liu14b}), the contrast in a signal must at least be larger than the noise level of a telescope in order to produce images. In this section we compare the typical contrast of the 21-cm signal predicted by our model (which we take to be the standard deviation of the one-point PDF) to the predicted noise levels from telescopes of comparable size to the full HERA array and the first phase of SKA-Low.

The brightness temperature of a region at redshift $z$, assuming it is fully heated and ignoring velocity fluctuations, is
\begin{equation}\label{eq:21cm_strength}
\delta T_b \approx 9 x_{\rm HI} (1 + \delta_R) \sqrt{1+z} \ \mbox{mK} ,
\end{equation}
which follows from eq. 18 in \cite{FOB06} with all terms in brackets taken to be unity. Here $\delta_R$ is the fractional overdensity of the region. On smoothing scales sufficiently small relative to the size of ionized bubbles, the one-point PDF will be dominated by a peak of fully ionized and one of fully neutral regions. The sensitivity needed for direct imaging of the 21-cm signal is thus often quoted as simply the contrast between these: $\sim 24[(1+z)/7]^{1/2}$~mK. Given that our model finds significantly smaller ionized bubble sizes than previous models (see section \ref{s:one-point-PDF-description}), the relevant smoothing scales are no longer much smaller than the typical ionized bubble, and so this estimate is no longer appropriate. Given a more careful treatment of the signal contrast from our model in combination with the significant reduction in ionized bubble size from the MFP, we find a smaller average contrast in the 21-cm signal than many previous estimates, indicating that higher sensitivities will be required for direct imaging of the epoch. We step through this in detail in the following sections, first using a simple analytic approximation and then results from our model.

\subsection{Simple estimate}\label{ss:simple_estimate}

To gain some intuition, let us refer to a very simple model. First, we imagine imaging on sufficiently large scales such that the mean free path is much smaller than the telescope resolution. In that case, we can assume that ionizing photons are not shared between telescope resolution elements. Second, we will approximate the galaxies as a single population with average number density $\bar{n}_g$, mass $m_g$, and ionizing efficiency $\bar{\zeta}$. We will also assume they have a linear bias $b_g$, so the local density of galaxies in a large region with density $\delta$ is $n_g = \bar{n}_g ( 1 + b_g \delta)$.
Then the average global ionized fraction is $\bar{x}_{\rm HII} = \bar{\zeta} (\bar{n}_g m_g / \bar{\rho}) = \bar{\zeta} f_{\rm coll}$,
where $\bar{\rho}$ is the average mass density of the Universe. Finally, we assume that the typical dispersion in the ionized fraction between resolution elements is $\ll 1$. This allows us to expand the ionization fraction term perturbatively. It is \emph{not} valid if a large fraction of regions on the relevant scale are fully neutral or fully ionized.

With our approximations, using eq. \ref{eq:21cm_strength}, the ionized fraction of a region is
\begin{equation}
x_{\rm HII} = \bar{\zeta} (\bar{n}_g m_g / \bar{\rho}) \approx \bar{\zeta}(\bar{n}_g m_g / \bar{\rho}) (1 + b_g \delta_R) \approx \bar{x}_{\rm HII} (1 + b_g \delta_R).
\end{equation}
Thus the difference between a patch and the average signal is
\begin{equation}
\delta T_b - \left< \delta T_b \right> \approx 9 \sqrt{1+z} [ 1 - \bar{x}_{\rm HI} (b_g+1)/b_g ] b_g \delta_R \ \mbox{mK}.
\end{equation}
Note that the factor in square brackets \emph{suppresses} the fluctuations. It arises because overdense regions, which naively have more signal because they have extra gas, are over-ionized, which removes some of that neutral gas. In other words, inside out reionization models tend to decrease the fluctuations relative to the matter power spectrum. 

The underlying density field is a gaussian, with standard deviation $\sigma_R$ = (0.048, 0.026, 0.016) for spheres of radius $R=25,\,50,\,75$~cMpc at $z=7$. The galaxy bias is $b_g = 3$--6.5 for halos from $M=10^{8}$--$10^{10} \ M_\odot$. Thus we would expect
\begin{equation}
\delta T_b - \left< \delta T_b \right> \approx 1.8 \left[ \frac{1 - \bar{x}_{\rm HI} (b_g+1)/b_g }{1/3} \right] \left[ \frac{b_g}{3} \right] \left[ \frac{ \sigma_R}{0.072} \right] \ \mbox{mK},
\end{equation}
where the factor of $1/3$ is appropriate for $\bar{x}_{\rm HI} = 0.5$ and $b_g = 3$. This simple model thus estimates the typical contrast in the signal $\mathcal{O}$(1 mK), an order of magnitude below that given by using the contrast between fully ionized and fully neutral regions.

\subsection{Sensitivity estimate using the full model}

We now return to our model for a more thorough treatment of the 21-cm signal contrast. Because the neutral fraction decreases with increasing density, the 21-cm signal is a non-monotonic function of the density (or neutral fraction), and there are exactly two densities (or neutral fractions) which correspond to each signal value. Thus, following \citetalias{FZH04}, the one-point PDF of the 21-cm signal is given by

\begin{equation} \label{eq:21cm}
    P_{21}(\delta T_b) = \sum_{i=1}^2 P_0(\delta_i) \left|{\frac{d \delta}{d (\delta T_b)}}\right|_{\delta_i},
\end{equation}
where the sum is over both densities $\delta_i$ which correspond to the same $\delta T_b$, and the derivative is evaluated at the appropriate $\delta_i$. In order to better capture the sub-structure of each smoothed region, we compute the 21-cm signal using the neutral fraction and density on the subregion scale ($R_{\rm sub} = 1$ cMpc in this work; see section \ref{ss:subregions}) using eq. \ref{eq:21cm_strength}. This follows \texttt{21cmFAST}, where the values on the voxel scale are used. As our model naturally outputs both the density and corresponding ionized fraction distributions on the subregion scale for smoothing scales of different densities, the extension to the 21-cm signal follows without any other modifications. Note that the non-monotonicity of the function means that the PDF becomes singular (though integrable) at a maximum signal strength where $\delta T_b(\delta)$ turns over.

\begin{figure}
    \centering
    \includegraphics[width=0.82\linewidth]{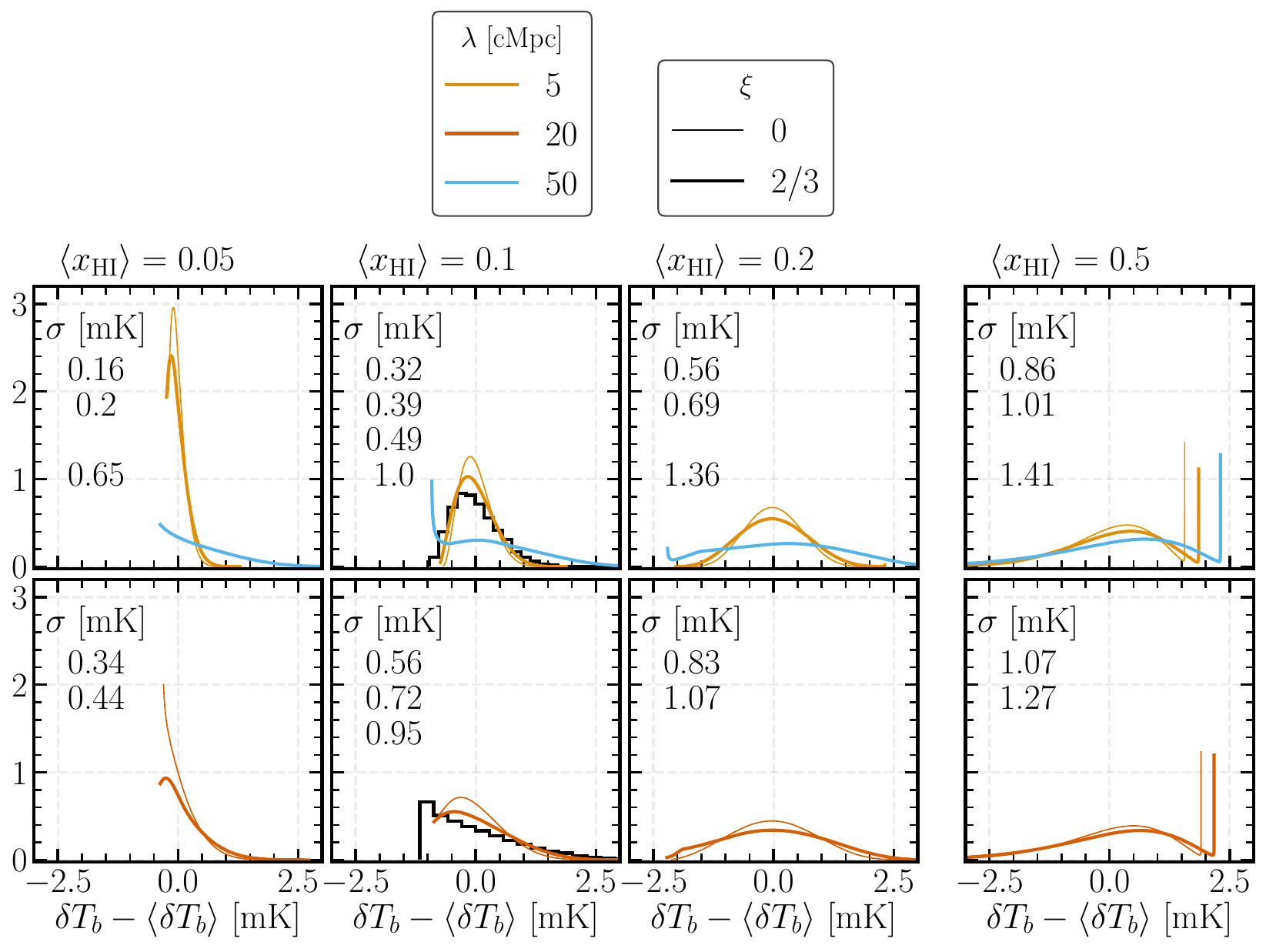}
    \caption{\textbf{New measurements of a small MFP at $z=6$ indicate that more sensitivity is required for direct images of the 21-cm signal compared with previous predictions.} Distributions show PDFs of the 21-cm signal strength for several different choices of the global ionized fraction (columns; the left three columns are computed at $z=6$ and the rightmost at $z=7$). All cases use a smoothing scale of $R=25$ cMpc. The mean 21-cm signal strength is subtracted in each case. Rows (and colors) show different choices for the MFP and linewidths indicate different mass-dependences of the ionizing efficiency (where $\xi=2/3$ uses the bursty version of the model). For comparison, we also include histograms of the result from \texttt{21cmFASTv1} in black for the $\left< x_{\rm HI} \right> = 0.1$ and $\xi = 0$ case as well as the result from our model using a low-$z$ extrapolation of the MFP ($\lambda=50$ cMpc using $\xi=2/3$, bursty; blue lines). In each panel from top to bottom, the standard deviation $\sigma$ is listed for $\xi = 0$, $2/3$, \texttt{21cmFAST}, and $\lambda = 50$ cMpc, where appropriate. In each case the PDF becomes singular at the maximum signal strength (see text for details), which does not always manifest as a peak in the plotted PDFs simply due to the finite step size used in the plotting software. The widths of these distributions indicate that sensitivities on the order of a mK or less are required for direct imaging of the 21-cm signal, much smaller than the naive estimate, and up to a factor of several smaller than is expected from the low-$z$ extrapolation of the MFP.}
    \label{fig:PTb_PDFs}
\end{figure}

In figure \ref{fig:PTb_PDFs} we show the one-point PDFs of the 21-cm signal at $z=6$ at various global neutral fractions using a smoothing scale of $R=25$ cMpc and $\lambda = 5$ and 20 cMpc. We find that the typical contrast of the signal is $\mathcal{O}$(1 mK) or less, consistent with our simple estimate in section \ref{ss:simple_estimate}, even when using an MFP value $> 3\sigma$ greater than the mean value measured by \citetalias{B21}. Even at the midpoint of reionization, where the signal contrast should reach a maximum, the variance barely surpasses 1 mK. For comparison, we also compute PDFs using \texttt{21cmFASTv1} with the same modified barrier which includes the MFP (\citetalias{DF22}) for $\left< x_{\rm HI} \right> = 0.1$ and $\xi=0$. Although the \texttt{21cmFASTv1} modification produces slightly wider distributions, they maintain a width of the same order as our analytic predictions. We note that \texttt{21cmFAST} accounts for velocity fluctuations when computing the 21-cm signal strength, which serves to widen the distributions compared with our method. We also show results using $\lambda = 50$ cMpc, approximately the value of a low-$z$ extrapolation of the MFP (\citetalias{B21}) and find distributions up to a factor of a few wider than using the mean value of $\lambda = 5$ cMpc, though still much narrower than the naive approximation would indicate. Previous works have found similar contrasts to this larger MFP case when computing the PDF on comparable scales to those considered in this work \cite{Mellema06, Ichikawa10, Kubota16}. We find that the short MFP decreases the contrast even further, up to a factor of several at lower global neutral fraction. Accounting for recombinations through a clumping factor, \cite{Wyithe07} found similarly narrow PDFs to our results which do include a short MFP, while \cite{Watkinson15} explored the effect of inhomogeneous sinks on one-point statistics using \texttt{21cmFAST} and found a similarly drastic reduction in the variance when compared with a model that contained no sinks.

\begin{figure}
    \centering
    \includegraphics[width=0.8\linewidth]{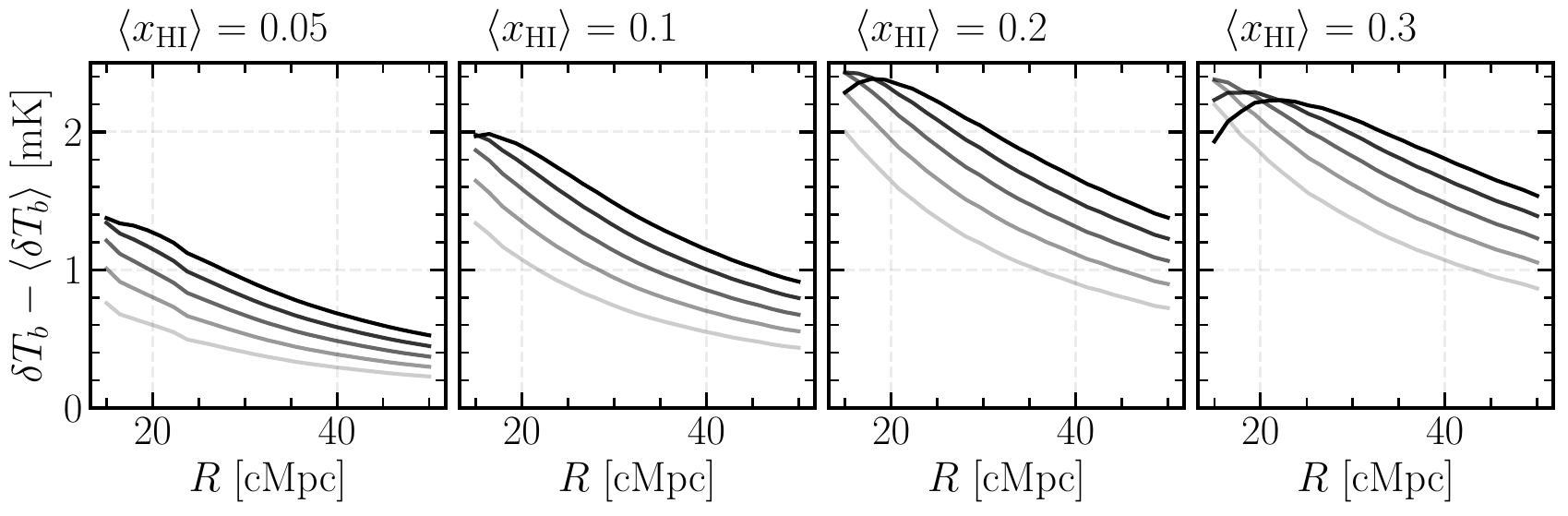}
    \caption{\textbf{Even the most extreme neutral regions have a low signal contrast.} Curves show the value of the 21cm signal, relative to the mean, of highly neutral regions as a function of smoothing radius. From light to dark (bottom to top in most cases), values are computed for neutral fractions corresponding to values of the CDF $P(> x_{\rm HI}) = 1/2 - {\rm erf}(n_\sigma/\sqrt{2})$, with $n_\sigma =$ 2, 2.5, 3, 3.5, and 4, equivalent to 2$\sigma$--4$\sigma$ for a Gaussian distribution. In most cases a more neutral region corresponds to a larger signal strength, but in some cases (particularly at higher global neutral fractions and smaller smoothing scales), less neutral regions actually have a higher signal strength. This is due to the non-monotonic nature of the 21cm signal (see eq. \ref{eq:21cm_strength}). All cases use $\xi=2/3$, the bursty version of the model, and $\lambda=5$ cMpc.}
    \label{fig:PTb_CDFs}
\end{figure}

As it is the neutral regions which actually produce the 21-cm signal, it is of interest to focus some attention on the most extreme neutral regions. To explore such regions, in figure \ref{fig:PTb_CDFs} we plot the value of the 21-cm signal, relative to the mean, of highly neutral regions as a function of the smoothing scale. We show results for regions with neutral fractions corresponding to values of the CDF $P(> x_{\rm HI}) = 1/2 - {\rm erf}(n_\sigma/\sqrt{2})$, with $n_\sigma =$ 2, 2.5, 3, 3.5, and 4 from top to bottom (light to dark), equivalent to 2$\sigma$--4$\sigma$ for a Gaussian distribution. Again, we find that even the most extreme neutral regions have $\delta T_b - \left< \delta T_b \right> \sim$ a few mK. In most cases, regions with more extreme neutral fractions have higher signal strengths. However, due to the non-monotonic nature of the 21-cm signal as a function of the neutral fraction, in some cases very high neutral fraction regions actually have a lower signal strength, particularly at higher global neutral fractions.

Thus, using the full results of our model, we find a typical contrast which is much lower than simple estimates and find that the shorter MFP reduces the contrast even further by up to a factor of several. In the following section we compare our signal predictions to noise estimates for HERA-like and SKA-like arrays to explore the implications of the short MFP on the observability of the 21-cm signal.

\subsection{Observational Prospects for HERA and SKA}

Ignoring foreground effects, we estimate the thermal noise limit for two idealized telescopes of comparable size to the full HERA array and the first phase of SKA-Low (SKA1-Low) using the standard radiometer equation including 
the array filling factor $\eta_f$ (e.g., \cite{FOB06}):
\begin{equation}\label{eq:noise}
    {\rm noise} = \frac{T_{\rm sys}}{\eta_f \sqrt{\Delta \nu \, t_{\rm int}}},
\end{equation}
where $T_{\rm sys}$ is the system temperature, $\Delta \nu$ is the bandwidth, and $t_{\rm int}$ is the integration time. $\eta_f = A/\pi r^2$, where $A$ is the total collecting area ($\sim 50,000\,$m$^2$ and $\sim 419,000\,$m$^2$ for each telescope, respectively) and $r$ is the baseline, related to the angular resolution and observed (i.e., redshifted) wavelength through $\theta = \lambda_{21,z}/r$. The system temperature is the sum of the receiver and sky temperatures: $T_{\rm sys} = T_{\rm rec} + T_{\rm sky}$, where the sky temperature is taken to be that of a quiet portion of the sky, $T_{\rm sky} = 180(\nu/180 \; {\rm MHz})^{-2.6}$ K. For the HERA analog, we take $T_{\rm rec} = 80$ K \citep{Kim25}, and for the SKA analog we assume the same.

\begin{figure}
    \centering
    \includegraphics[width=0.7\linewidth]{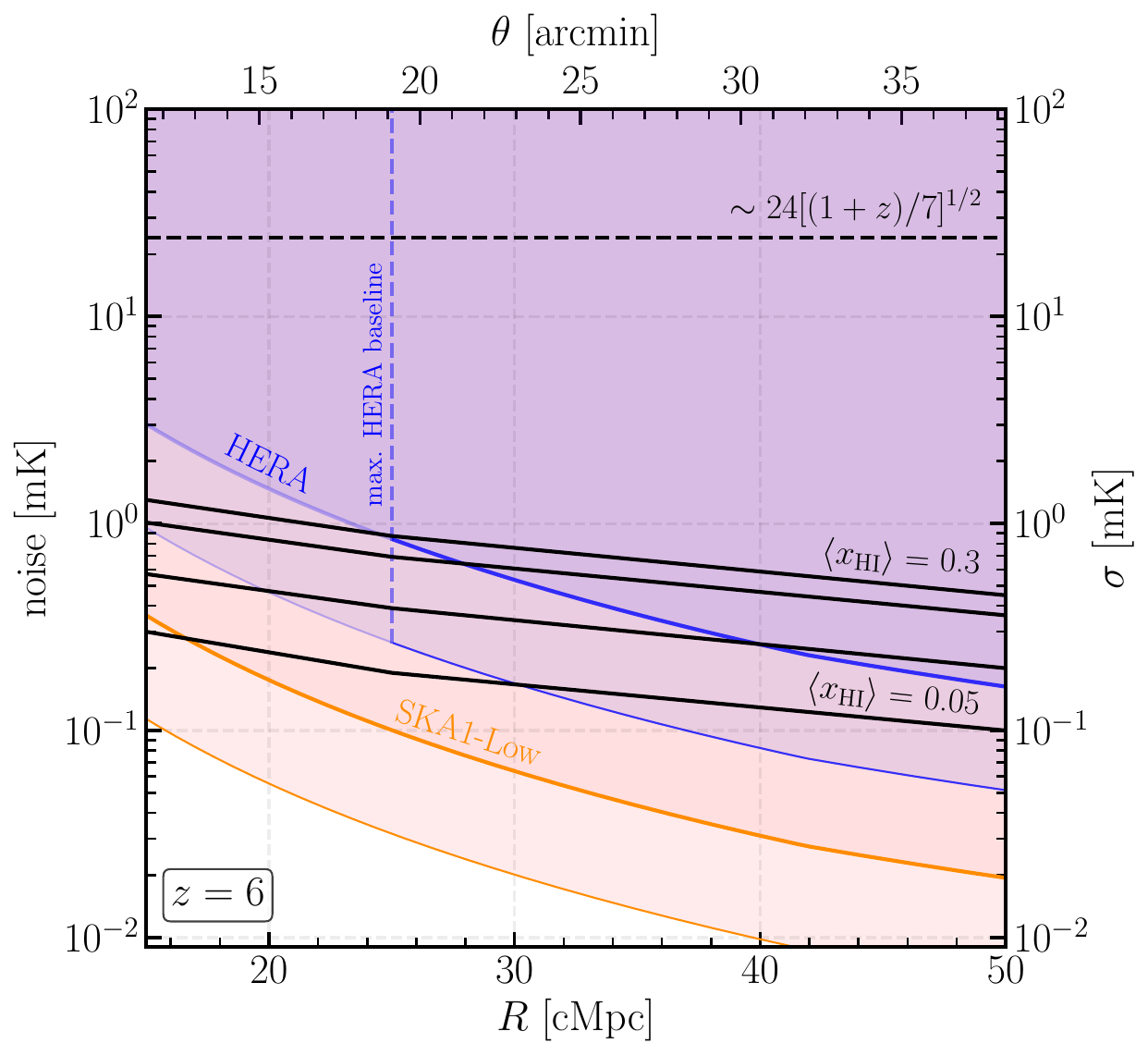}
    \caption{\textbf{New predictions for the 21-cm signal strength given a shorter MFP push up against the limits of the capabilities of upcoming radio interferometers.} Solid colored lines show the estimated noise levels for idealized telescopes with similar sizes to HERA (blue) and SKA1-Low (orange) for integration times of 100 hr (thick lines) and 1000 hr (thin lines) at $z=6$. We show results across a range of resolutions, corresponding to regions with $R = 15$--50 cMpc at this redshift. Note that we associate the angular resolution with the diameter of the smoothed region. The vertical dashed blue line shows the approximate maximum baseline for an array of the same size as the full HERA core. Solid black lines show our predictions for the standard deviations of the one-point PDF of the 21-cm signal at global ionized fractions of $\left< x_{\rm HI} \right> = 0.05,0.1,0.2,$ and 0.3 from bottom to top, with $\lambda = 5$ cMpc and $\xi=2/3$, bursty. The dashed black line shows the point of comparison often quoted for sensitivity predictions --- the contrast between fully ionized and fully neutral regions --- which is valid only when $R$ is much smaller than the typical ionized bubble. Given a more careful calculation in combination with a shorter MFP, we find regions with a contrast 1--2 orders of magnitude smaller. At the largest scales shown the widths of the 21-cm distributions are an order of magnitude or more larger than the predicted noise levels of the SKA analog, but only a factor of a few larger than those of the HERA analog, or even smaller in some cases.}
    \label{fig:sensitivity}
\end{figure}

In figure \ref{fig:sensitivity} we show the noise levels for both telescopes for integration times of $t_{\rm int} = 100$ and 1000 hr at resolutions corresponding to $R = 15$--50 cMpc at $z=6$. At each resolution the bandwidth $\Delta \nu$ corresponds to the same comoving distance as the angular scale. In addition, we plot the standard deviations of the 21-cm one point PDFs computed using our model ($\lambda=5$ cMpc, $\xi=2/3$, bursty) across the same scales at various global neutral fractions. We also show the contrast between fully ionized and fully neutral regions ($\sim 24 [(1+z)/7]^{1/2}$). At these integration times, both telescopes have enough sensitivity to comfortably image signals at this larger contrast. However, given the shorter MFP and a more careful treatment of the one-point PDF, we find regions with a signal contrast 1--2 orders of magnitude smaller. While the noise levels of the SKA-Low analog fall an order of magnitude or more below the signal widths on the largest scales shown here, our predictions for the signal strength push up against the limits of the capabilities of the HERA analog, falling only a factor of a few above --- or even below --- the predicted noise levels, depending on the integration time and global ionized fraction. Note that the HERA core has a maximum baseline of $\sim 300$ m, corresponding to a maximum resolution of $\sim 19$ arcmin or a minimum radius of $R \sim 25 $ cMpc. Outriggers increase the maximum baseline beyond this scale but can no longer be represented by a filled aperture and are therefore not well approximated by eq. \ref{eq:noise}.

We note that these noise levels and indeed the comparison of noise level to one-point PDF width are only approximate. However, it is clear that significantly more sensitivity is required for 21-cm tomography than is expected from the common simple contrast estimate, something that has also been found in other careful considerations of the one-points statistics of the 21-cm signal. While there have been some detailed studies of telescope noise and other observational effects in this context \citep{Kittiwisit18,Kittiwisit22,Kim25}, we find that the short MFP exacerbates this issue, reducing the contrast even further, indicating that it should be considered in sensitivity predictions or statistical constraints of the EoR.

\section{Conclusion} \label{s:conclusion}
In this work we describe the first analytic model for reionization which explicitly incorporates the MFP as a free parameter. Our model is based on the excursion set theory, which allows us to describe the ionization field through the statistics of the density field in combination with an ionization criterion which dictates the minimum density on each scale for a central subregion to be ionized. \citetalias{DF22} determined such a criterion which explicitly incorporates the MFP of ionizing photons and applied this to the semi-numerical simulation \texttt{21cmFAST}, which we modify to include a mass-dependent ionizing efficiency. By applying this barrier to an analytic model we are able to extract the first-crossing distribution and therefore the size distribution of ionized bubbles. With an eye towards direct imaging of the 21-cm signal, we extend the model to describe partially ionized regions by including a statistical description of subregions --- and therefore substructure --- in each larger region, which allows us to extract the one-point PDF of the ionization field. Through a simple association of density and ionized fraction we then extend the model to describe the one-point PDF of the 21-cm signal, the observable quantity in this case.

We find that the MFP has a significant effect on the topology of reionization, and that the effect is much more pronounced towards the end stages of reionization, where near-future observations are most relevant. Qualitatively, this is easily seen in the ionization boxes output by \texttt{21cmFAST}. Quantitatively, we explore the change in topology through the size distribution of ionized bubbles and the one-point PDF of the ionization field. Specifically, we find that the MFP values measured by \citetalias{B21} result in ionized bubble sizes close to an order of magnitude smaller than models which ignore the MFP, and bubbles which are still smaller than models which approximate the MFP with a strict upper cutoff radius. We also find that the one-point PDF is sensitive to the particular value of the MFP (and other parameters) at lower $\left< x_{\rm HI} \right>$, and thus conclude that the PDF (or CDF) remains a potentially useful observational probe of reionization models.

We find that the resulting 21-cm signal contrast is an order of magnitude or more below a commonly-used estimate, owing to the significant reduction in ionized bubble sizes caused by a shorter MFP in combination with a more thorough calculation of the contrast itself. While existing studies have found a similar reduction in contrast without a shorter MFP, we find that the shorter MFP further decreases the contrast by up to a factor of a few. We compare the predicted contrast to noise levels for upcoming radio interferometers and find that the necessary sensitivity pushes up against the predicted limits of such telescopes, indicating that more detailed studies of predicted noise levels with respect to a reduced signal strength --- including the effect of a short MFP --- are needed.

Our results indicate that the effect of such a short MFP on reionization is significant, warranting careful treatment of the parameter in the future development of models of the epoch. Our model, although approximate in many ways, allows an efficient and physically transparent method with which to explore the effects of this parameter on large scales for a range of observables.

\acknowledgments
We thank Jun Zhang, Zhilei Xu, Andrei Mesinger, Jonathan Pober, and Sahil Hegde for helpful discussions. This work was supported by NASA through award 80NSSC22K0818 and by the National Science Foundation through award  AST-2205900. In addition, this work was supported in part by the Queen's Road Foundation. This material is also based upon work supported by the National Science Foundation under grant nos. AST-1636646 and AST-1836019 and institutional support from the HERA collaboration partners. This work has made extensive use of the SAO/NASA Astrophysics Data System (\href{http://ui.adsabs.harvard.edu}{ui.adsabs.harvard.edu}) and
Science Explorer (\href{https://scixplorer.org}{scixplorer.org}), the arXiv e-Print service (\href{http://arxiv.org} {arxiv.org}), as well as the following softwares: \texttt{matplotlib} \cite{Matplotlib}, \texttt{numpy} \cite{numpy}, \texttt{astropy} \cite{Astropy}, and \texttt{scipy} \cite{Scipy}.

\appendix

\section{Comparison with 21cmFAST}\label{a:21cmFAST-differences}

In this section we describe several important differences between our model and \texttt{21cmFAST} which may contribute to discrepancies in some results. Specifically, we compare only to a modified version of \texttt{21cmFASTv1} which computes the ionization field directly from the overdensity field using the same ionization criterion as our model. Because \texttt{21cmFAST} generates 3D realizations of the density field (and other fields), more complex treatments are also possible. Some such treatments are explored in \citetalias{DF22} and we also point the reader to \cite{Davies25} for a full description of the most recent public release of \texttt{21cmFAST} which includes additional detailed physics. In this sense, the discussion here highlights only the discrepancies that occur because our model lacks some information that is contained within a density field with 3D spatial correlations, rather than any additional details which may be present in newer versions of \texttt{21cmFAST} or other semi-numerical models.

Firstly, the mapping from density to ionized fraction in our model is one-to-one. One can think of the model as having an infinite number of subregions per region (see section \ref{s:one-point-PDF-description}), which has the effect of requiring a region to be at or above the barrier in order to be fully ionized, as the only way that all random walks associated with a region reach the barrier is if the region itself has a density at or above the barrier. Note that this is not an enforced criterion, but instead arises naturally from the description in section \ref{s:one-point-PDF-description}. For a fixed $\zeta_0$, the barrier defined by eq. \ref{eq:ion-bar} is always smaller for larger values of $\lambda$, meaning that random walks are more likely to cross at all radii. However, in order to achieve the same global ionized fraction, $\zeta_0$ must be larger for smaller values of the MFP, meaning that in many cases the barriers for two different MFP values intersect, and become \textit{smaller} for smaller values of $\lambda$ at small $R$. Because the fraction of fully ionized regions in our model is determined solely by the density on the smoothing scale, this leads to more ionized regions at small radii for smaller MFP values, causing the turnover.

In \texttt{21cmFAST} there are of course a finite number of subregions (voxels) per region ($n_{\rm sub} \sim R^3/R_{\rm sub}^3$). This leads to a scatter between the density of a region and its ionized fraction, and is more significant at smaller smoothing scales as $n_{\rm sub}$ becomes small. Although this will change the one-point PDF (through a convolution of the scatter in the ionized fraction of each density at each scale), it will have a more dramatic effect on fully ionized regions, as many up-scattered regions will ``bunch up'' as fully ionized. We can include this scatter by reconstructing our model in a Monte Carlo manner and only considering $n_{\rm sub}$ subregions per smoothed region. We show a comparison of all three models in figure \ref{fig:PQ1_funcR}. At large scales ($R \gtrsim 15$ cMpc), all models more or less agree, as $n_{\rm sub}$ is sufficiently large. At small scales, as expected, \texttt{21cmFAST} exceeds the analytic case. While the Monte Carlo version of the model does increase the fraction of fully ionized regions at small scales compared with the analytic model, in most cases it still does not reach the fractions in \texttt{21cmFAST}. This is likely because this model also ignores another source of scatter present in \texttt{21cmFAST}. 

A second source of scatter arises from the approximation in the analytic excursion set that all random walks associated with subregions in a given smoothed region pass through the density of the smoothed region on the smoothing scale. In \texttt{21cmFAST}, the random walks associated with adjacent pixels which constitute a single smoothed region all pass through different densities on the smoothing scale. Imagine a situation where, for a small radius, a smoothed region's density exceeds the $\lambda = 5$ cMpc barrier but not that of $\lambda = 20$ cMpc. Without scatter, this leads to the turnover where more regions are ionized for the shorter MFP value. This effect will scatter some of the densities above the $\lambda = 20$ cMpc barrier and below the $\lambda = 5$ cMpc barrier, decreasing this discrepancy. However, in this case there are still more regions above the shorter MFP barrier. However, in the case of a finite number of subregions, we must also consider the behavior of random walks on other scales. At larger radii, where the $\lambda = 5$ cMpc exceeds that of the longer MFP value, more random walks cross the latter. These effects combine, leading to more overall ionized regions on the smaller scale.

Besides these discrepancies which are, as mentioned above, more severe for fully ionized regions, our model generally agrees with PDFs created using \texttt{21cmFAST} (\texttt{v1} with the same barrier), indicating that the method of subregions accounts for much of the spatial information necessary for computing the one-point PDF, despite not using a fully realized 3D density field.

\section{Effect of finite simulation boxes on the one-point PDF} \label{s:finite_box}
Many simulations of reionization fix the mean ionized fraction of their simulation boxes to the desired global ionized fraction. If the simulation volume is too small, this can affect the statistics of various quantities in the simulation \citep{Barkana04}. Because the maximum scale of our model is arbitrary, we can use it to easily explore the effect of this finite box size on the one-point PDF. Before proceeding, we clarify that the results of our model are not sensitive to cosmic variance. That is, the model is not sensitive to the effect of a limited number of sampled regions when computing the one-point PDF. This will allow us to easily isolate the effect of a limited simulation box on the ionization topology itself from the statistical effects of cosmic variance on the sample.

\begin{figure}
    \centering
    \includegraphics[width=1.0\linewidth]{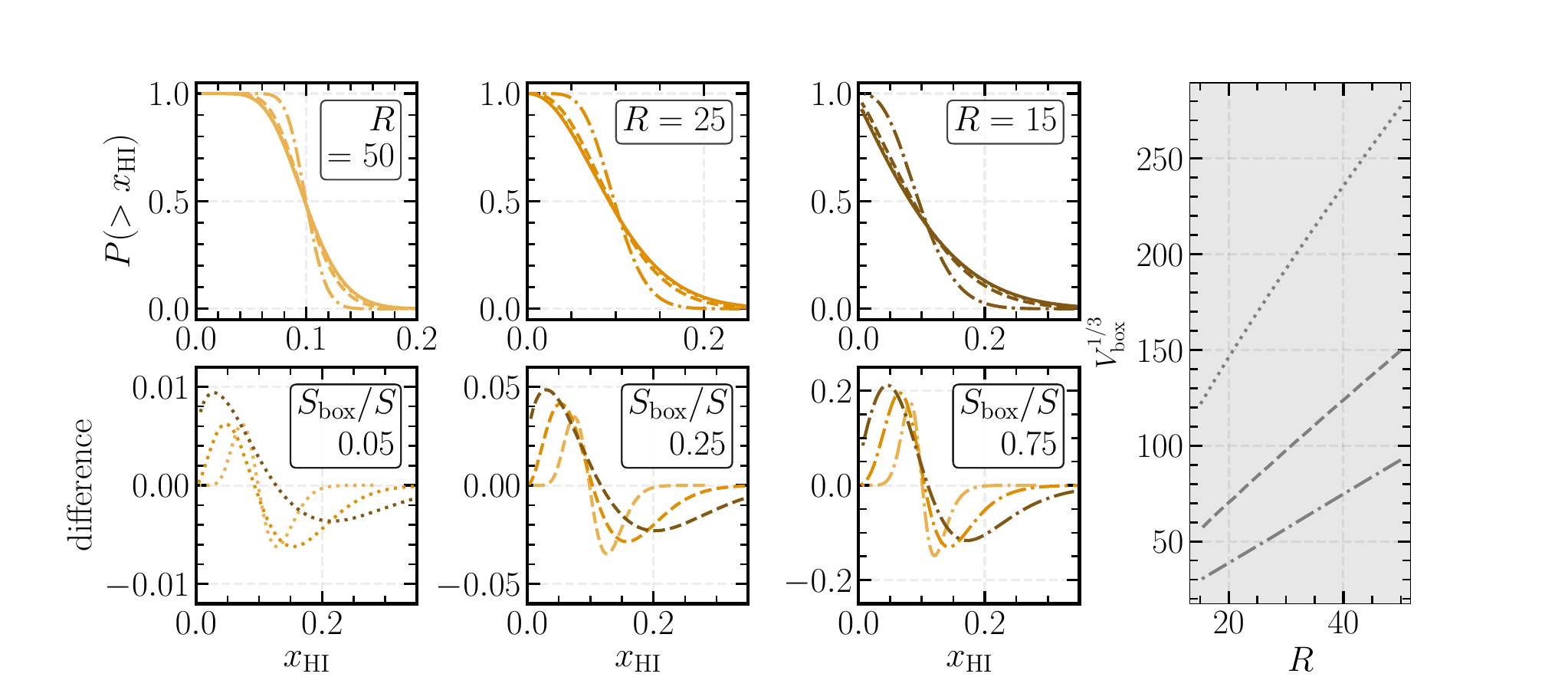}
    \caption{\textbf{The maximum difference in the CDF due to a finite simulation box is approximately a function of the ratio of the variance of the box to that of smoothing scale, $S_{\rm box}/S$, independent of the absolute value of smoothing scale.} Each panel in the top row shows a comparison of CDFs evaluated at the same smoothing scale (colors) but different ``box sizes'' (which is represented in our model by a finite starting scale; linestyles). All cases use $\lambda=5$ cMpc and $R = 25$ cMpc. We show results for the ``true'' distribution (which uses a very large box size with $R_{\rm box} = 1000$ cMpc, or $S_{\rm box}/S \sim 0$; solid lines), and various ratios of the variance of the box to that of the smoothing scale: $S_{\rm box}/S \sim 0.05$ (dotted lines), $S_{\rm box}/S \sim 0.25$ (dashed lines), and $S_{\rm box}/S \sim 0.75$ (dot-dashed lines). Dotted lines are present in the top row, but are so close to the solid lines as to not be visible. Note that larger ratios of $S_{\rm box}/S$ means a \textit{smaller} box size, as increasing $R$ corresponds to decreasing $S$. Bottom panels show the difference between CDFs computed using finite box sizes to the ``true'' distribution. In this case, each panel corresponds to the same value of $S_{\rm box}/S$, but different smoothing scales. We find that the maximum magnitude of the deviation from the true distribution caused by a finite box size is approximately constant with $S_{\rm box}/S$, independent of the absolute value of the smoothing scale. However, there is a slight trend of a larger deviation for smaller absolute smoothing scales. For reference, the right panel shows the box side length ($V_{\rm box}^{1/3}$) required to achieve each value of $S_{\rm box}/S$, as a function of smoothing scale. For example, to ensure a deviation no larger than $\sim 0.01$ requires $S_{\rm box}/S \sim 0.05$, and so for a smoothing scale of 50 cMpc a box of side length $\sim 280$ cMpc is required. However, to ensure the same deviation for a smoothing scale of 25 cMpc, a box of side length of only $\sim 170$ cMpc is required.}
    \label{fig:CDF_diff_Rbox}
\end{figure}

We can easily replicate the effect of a finite box by changing the starting scale of the excursion set in our model. Rather than starting the random walks at $(S,\delta) \sim (0,0)$ as is assumed in eq. \ref{eq:cross_smallS}, we can instead start them at $(S,\delta) = (S_{\rm box},0)$ (or indeed any arbitrary starting point).\footnote{In practice, because we sample $Q(\delta|S)$ for a finite number of densities on a given scale $S$, it is also beneficial to alter the range of sampled densities by accounting for the reduced variance over the random walk scales considered, as the probability of densities in this case is no longer given by $P_0(\delta',S)$, but instead $P_0(\delta',S-S_{\rm box})$.} Broadly, we explore this effect by first computing the ``true'' distribution using a very large starting radius and compare this to distributions computed using various starting (or box) scales. As is done throughout this work, we take $R = 1000$ cMpc ($S \sim 10^{-5}$) as our very large starting radius. Because the random walks are sensitive to the relative difference between the beginning variance scale $S_{\rm box}$ and the smoothing scale $S$, rather than compare to starting points of equal radii, we instead compare to those with equal ratios of the starting scale to the smoothing scale, $S_{\rm box}/S$. Our true distribution has $S_{\rm box}/S \sim 0$, and we compare to boxes with $S_{\rm box}/S = 0.05, 0.25,$ and 0.75 in figure \ref{fig:CDF_diff_Rbox}. Each panel in the top row shows the CDFs at the same smoothing radius $R$ for different values of $S_{\rm box}/S$, represented by line styles. As expected, there is a larger discrepancy for larger values of $S_{\rm box}/S$ (smaller box sizes). To quantify the discrepancy, bottom panels show the difference between CDFs using finite box sizes to the ``true'' distribution for constant values of $S_{\rm box}/S$ at each of the three smoothing radii shown. From this we can see that the maximum difference between the CDF resulting from a finite box size is dependent only on the ratio $S_{\rm box}/S$, independent of the absolute smoothing scale. For example, the maximum difference from the true CDF is $~0.01$ if $S_{\rm box}/S = 0.05$, regardless of the smoothing radius used. For reference, the right-hand panel shows the box sidelength necessary, as a function of smoothing radius, to achieve the three values of  $S_{\rm box}/S$ shown. For example, to ensure a deviation no larger than $\sim 0.01$ requires $S_{\rm box}/S \sim 0.05$, and so for a smoothing scale of 50 cMpc a box of side length $\sim 280$ cMpc is required. However, to ensure the same deviation for a smoothing scale of 25 cMpc, a box of side length of only $\sim 170$ cMpc is required. Such box sizes are moderate and are certainly achievable by some state-of-the-art semi-numerical and even numerical simulations. We checked the same distributions using $\lambda=20$ cMpc and find very nearly the same result.

\bibliography{draft_final.bib}{}
\bibliographystyle{JHEP.bst}

\end{document}